\documentstyle[preprint,aps,psfig]{revtex}

\def\wtilde{\tilde {\omega }}

\def\YBCO{YBaCu$_3$O$_{7-x}$}

\def\bea {\begin{eqnarray}}
\def\eea {\end{eqnarray}}
\def\be {\begin{equation}}
\def\ee {\end{equation}}
\def\bff{ }

\begin{document}
%\markboth{PHYSICAL REVIEW B {\bf 52}, 123 (1996)}{cond-mat/9709242}
\draft

\title
{Quasiparticle Lifetimes and the Conductivity Scattering Rate}
\author{F. Marsiglio$^{1,2,3 \dagger}$ and J.P. Carbotte$^{2,3}$}
\address
{$^1$Neutron \& Condensed Matter Science\\
AECL, Chalk River Laboratories, Chalk River, Ontario, Canada K0J 1J0\\
$^2$Dept. of Physics \& Astronomy, McMaster University,
Hamilton, Ontario L8S 4M1 \\
$^3 $ Canadian Institute for Advanced Research, McMaster University,
Hamilton, ON L8S 4M1}
\date{June 24, 1997}
% \today} 
% Dec. 3, 1996}

% This is the secret. The closing ``]'' on the line below...
%\twocolumn[\hsize\textwidth\columnwidth\hsize\csname@twocolumnfalse\endcsname
\maketitle
\begin{abstract}
We compute the single-particle inverse lifetime, along with the
conductivity-derived scattering rate, for a metallic system
in an s-wave superconducting state.
When both electron-phonon and electron-impurity scattering are
included, we find that while these scattering rates are in
qualitative agreement, in general quantitative agreement is lacking.
We also derive results for the quasiparticle lifetime within the BCS
framework with impurity scattering, which makes it clear that impurity
scattering is suppressed for electrons near the Fermi surface in the
superconducting state.

%\\{\tt preprint: cond-mat/9709242}
\end{abstract}
%\pacs{\phantom{PACS numbers: 74.20.-z, 74.20.Fg, 74.25.Dw, 74.25.Nf}}
% ...is the rest of the secret.
%]

\section{INTRODUCTION}

The quasiparticle lifetime is a concept which is useful 
for clarifying the nature of 
the system of interest, i.e. is it a Fermi vs. marginal Fermi vs. a 
Luttinger liquid, etc (Bedell {\it et al.}, 1990).
Various techniques are available for measuring
lifetimes (Kaplan {\it et al.} 1976); perhaps the most direct is through tunnel
junction detection (Narayanamuti {\it et al.} 1978). In this paper we analyze
electron scattering rates,  as measured by microwave
and far-infrared conductivity measurements, and examine their relationship to
quasiparticle lifetimes. {\bff We will reserve the name
``quasiparticle lifetime''
to refer to the single-particle property to be defined technically in the next
section. In contrast we will use the term ``scattering rate'' or
``scattering lifetime'' to refer to a property derived from a response function,
such as the optical conductivity in this case.} The experimental results for
\YBCO (Bonn {\it et al.} 1992, 1993) and their theoretical implications
(Bonn {\it et al.} 1993; Berlinsky {\it et al.} 1993; Klein 1994;
Bonn {\it et al.} 1994) have been discussed extensively in 
the literature. {\bff In these works the microwave conductivity was used
to extract information about quasiparticle scattering; it is commonly and
often tacitly presumed that the ``lifetime'' (inverse scattering rate)
derived from these sorts of measurements is closely related to the
quasiparticle lifetime, as derived from the single-particle Green function.}
{\bff In this work we explore this relationship and compute both
the single-particle lifetime and the
two-particle scattering rate, including both electron-impurity scattering
and electron-phonon scattering. The latter involves inelastic scattering
processes and introduces significant complications. We use an Eliashberg
formalism to analyze this problem, with some further approximations which will
be made clear in the relevant sections.
A similar approach has already been used to analyze the normal state
problem (Shulga {\it et al.} 1991; Marsiglio and Carbotte 1995).
Here, certain aspects of quasiparticle lifetimes are clarified
and the analysis is extended to the superconducting state,
following our recent work on the
optical conductivity (Marsiglio and Carbotte 1995; Marsiglio {\it et al.}
1996).\par

We will proceed by carefully defining the single-particle lifetime.
We review two ways in which conductivity data can
be used to extract a scattering rate, one based on the two-fluid
model, and the other based on a straightforward Drude fit to the low 
frequency conductivity. The instances in which these procedures
yield a qualitative or quantitative facsimile of the quasiparticle inverse
lifetime will be clarified. In this way, one can evaluate the usefulness of the
two-fluid hypothesis, for example, in systems with both elastic and inelastic
scattering channels.\par

This paper is divided into
two sections. The first reviews quasiparticle lifetimes, which are 
calculated from the single-particle Green function. {\bff In the normal state
the single-particle equations for
electron-impurity scattering (in the Born approximation)
and electron-phonon scattering are well known
(Mahan 1981; Allen and Mitrovi\'{c} 1982; Grimvall 1981).
In the superconducting state, we first examine the BCS case with
electron-impurity scattering. By ``BCS'' we mean the limit where the
pairing interaction is instantaneous, and in this case, local, i.e. the
gap function is independent of frequency and momentum. The
single-particle spectral function is given analytically, showing explicitly
the role of the so-called BCS coherence factors which are somewhat disguised
in the clean limit. When impurity scattering is included, such an
expression should be used in lieu of the more
commonly adopted phenomenological form found in standard texts
(Schrieffer 1983).
When inelastic electron-phonon scattering is included in the problem,
perturbative methods within the Eliashberg framework are used. This follows
closely the work in Kaplan {\it et al.} (1976); however, here we discuss the
quasiparticle lifetimes in the presence of {\em both} impurity and
phonon scattering.}\par

The second section elucidates two methods recently
used to extract scattering rates from optical and microwave conductivity
measurements, and how these relate to the quasiparticle inverse lifetime.
We should make it clear at the start that the calculation we use for
the optical conductivity omits vertex corrections. The main result
of this is that the ``$1 - \cos{\theta}$'' factor that occurs in a
Boltzmann formulation of transport properties is absent (Mahan 1981),
so in fact, when theoretical comparisons between the quasiparticle
lifetime and the conductivity-derived scattering rate are made, the
agreement will in general be better than it would be, given an exact
calculation. Conversely, poor agreement between these
two quantities in the theory would almost certainly result in poor agreement
between the two measured quantities.

\section{QUASIPARTICLE LIFETIMES}

Quasiparticle lifetimes in clean electron-phonon superconductors were discussed
long ago by Kaplan {\it et al.} (1976). Before proceeding to their
relationship with the optical conductivity we give a brief review of the
formalism (Kaplan {\it et al.} 1976; Nicol 1991).
{\bff In the normal state the quasiparticle
has energy and lifetime defined by the pole of the single 
particle retarded Green function, i.e. the zero of
\begin{equation}
G^{-1}(k,\omega + i\delta) = \omega - \epsilon_k - \Sigma(\omega + i \delta),
\label{zero}
\end{equation}
\noindent where $\Sigma(\omega + i \delta)$ is the electron self-energy.
More specifically the solution $\omega_k$, where $\omega_k = \epsilon_k
+ \Sigma(\omega_k + i \delta)$, defines
the quasiparticle energy $E_k$ and inverse lifetime $\Gamma_k$ by
\begin{equation}
\omega_k \equiv E_k - i \Gamma_k/2
\label{pole}
\end{equation}
\noindent Similarly,} in the superconducting state, the inverse quasiparticle
lifetime is defined by (twice) the imaginary part of the pole in
the single-particle Green function. The
diagonal component of the single-particle Green function is
(Scalapino 1969)
\begin{equation}
G_{11}(k,\omega) = { \omega Z(\omega) + \epsilon_k \over
\omega^2 Z^2(\omega) - \epsilon_k^2 - \phi^2(\omega) },
\label{green}
\end{equation}
\noindent where
$Z(\omega)$ and $\phi(\omega)$ are the renormalization and pairing functions
given by solutions to the Eliashberg equations
(Eliashberg 1960; Scalapino 1969; Marsiglio {\it et al.} 1988)
which are repeated here for convenience:
\begin{eqnarray}
\phi(\omega) & = & \pi T\sum\limits_{m=-\infty}^{\infty}
\bigl[\lambda(\omega-i\omega_m)-\mu^*(\omega_c)\theta(\omega_c-\vert{\omega_m}
\vert)
\bigr]{\phi_m\over\sqrt{\omega^2_m Z^2(i\omega_m)+\phi_m^2}}
\nonumber \\
 & & +i\pi\int_0^{\infty}d\nu\,\alpha^2F(\nu)\Biggl\{\bigl[N(\nu)+f(\nu-
\omega)\bigr]{\phi(\omega-\nu)\over\sqrt{\wtilde^2(\omega-\nu) -
\phi^2(\omega-\nu)}}
\nonumber \\
 & & +\bigl[N(\nu)+f(\nu+\omega)\bigr]{\phi(\omega+\nu)\over\sqrt{
\wtilde^2(\omega+\nu) - \phi^2(\omega+\nu)}}\Biggr\}
\label{phireal}
\end{eqnarray}
\noindent and
\begin{eqnarray}
\wtilde(\omega) & = & \omega + i\pi T\sum\limits_{m=-\infty}^{\infty}
\lambda(\omega-i\omega_m)
{\omega_m Z(i\omega_m)\over\sqrt{\omega^2_m Z^2(i\omega_m)
+\phi_m^2}}
\nonumber \\
 & & + i\pi \int_0^{\infty}d\nu\,\alpha^2F(\nu)\Biggl\{\bigl[N
(\nu)+f(\nu-\omega)\bigr]{\wtilde(\omega-\nu)\over\sqrt{\wtilde^2(\omega-\nu)
-\phi^2(\omega-\nu)}}
\nonumber \\
& & +\bigl[N(\nu)+f(\nu+\omega)\bigr]{\wtilde(\omega+\nu)\over\sqrt{
\wtilde^2(\omega+\nu)-\phi^2(\omega+\nu)}}\Biggr\}\ ,
\label{wtildereal}
\end{eqnarray}
\noindent where $\wtilde(\omega) \equiv \omega Z(\omega)$.
Here, $N(\nu)$ and $f(\nu)$ are the Bose
and Fermi distribution functions, respectively. The electron-phonon spectral 
function is given by $\alpha^2F(\nu)$, its Hilbert transform is
$\lambda(z)$. The Coulomb repulsion parameter is $\mu^\ast(\omega_c)$
with cutoff $\omega_c$. A negative value for this parameter
can be used to model some BCS attraction of unspecified origin.
The renormalization and pairing
functions are first obtained on the imaginary axis at the 
Matsubara frequencies, i.e. $\omega = i\omega_n \equiv
i\pi T(2n-1)$, with $\phi_m \equiv \phi(i\omega_m)$ by setting the complex
variable $\omega$ in these equations to the Matsubara frequencies 
(Owen and Scalapino 1971; Rainer and Bergmann 1974).
Then the equations are iterated as written, with $\omega$ set to a frequency
on the real axis. Note that the square roots with complex arguments are
defined to have a positive imaginary part.\par

Actually, these functions are obtained at frequencies just above the real axis,
and then, in principle, the pole is given by the zero of the denominator 
continued to the lower half-plane. However, since the solutions are
readily known only along certain lines in the complex plane (e.g. the imaginary
axis or just above the real axis) we follow previous authors and look
for the pole perturbatively. That is, we write $\omega \equiv E - i\Gamma$,
and linearize the imaginary part so that (Scalapino 1969;
Kaplan {\it et al.} 1976)
\begin{equation}
\Gamma(E) = { EZ_2(E) \over Z_1(E)} - {\phi_1(E) \phi_2(E) \over EZ_1^2(E)}
\label{lifetime}
\end{equation}
\noindent Note
that $E$ is determined by equating the real parts (again linearizing 
in imaginary components), which yields
\begin{equation}
E = \sqrt{{\epsilon_k^2 + \phi_1^2(E) \over Z_1^2(E)}}.
\label{energy}
\end{equation}
{\bff In the normal state, $\phi(\omega) \equiv 0$, and Eq. (\ref{wtildereal})
simplifies to the known normal state result
(Allen and Mitrovi\'{c} 1982; Grimvall 1981).
The quasiparticle energy and lifetime are given by Eqs.(\ref{energy}) and
(\ref{lifetime}), which yield in the low energy limit
(Allen and Mitrovi\'{c} 1982; Grimvall 1981):
\begin{equation}
E \approx { \epsilon_k \over 
1 + \lambda^\ast(T) }
\label{enerk}
\end{equation}
\begin{equation}
\Gamma(E = 0) \approx { {1\over \tau} + 4\pi\int_0^\infty d\nu \alpha^2F(\nu)
\bigl[N(\nu) + f(\nu) \bigr] \over 1 + \lambda^\ast(T) }
\label{gammak}
\end{equation}
\noindent where
$f(\nu)$ is the Fermi function and
\begin{equation}
\lambda^\ast(T) \approx -{1\over \pi T} \int_0^\infty d\nu \alpha^2F(\nu)
Im \psi^\prime({1 \over 2} + i{\nu \over 2 \pi T}).
\label{lambdastar}
\end{equation}
Typical results for various electron-phonon
spectral functions are shown in Fig. 1 for the clean limit over a wide range
of temperatures.\par

Returning to the superconducting state}, at the Fermi surface
$\epsilon_k \equiv 0$ so Eq. (\ref{energy}) gives $E = \phi_1(E)/Z_1(E)
\equiv \Delta_1(E)$, which  becomes the definition for the lowest energy 
excitation, i.e. the gap in the excitation spectrum. It has become common
practice to consider $E$ as an independent variable, and then to study
$\Gamma(E)$ as a function of $E$. In Fig. 2 we show $\Gamma(E)$ vs. $E$
for various temperatures, using a Debye model spectrum. It is clear
that the scattering rate near the gap edge (shown by the arrow)
decreases very quickly as the temperature decreases. Note
that $\Gamma(E)$ actually becomes negative at intermediate temperatures
near $E \approx 10$ meV. This is real (i.e. not a numerical artifact)
and is simply a property of the function $\Gamma(E)$ given by 
Eq. (\ref{lifetime})
through the linearization procedure. The true pole in Eq. (\ref{green}) will
always
have a negative imaginary part (i.e. $\Gamma$ is positive).\par

What happens when both the BCS limit and the clean limit are taken ? 
Then $Z_2(E) \rightarrow 0$ and
$\phi_2(E) \rightarrow 0$, i.e. the functions involved in the solution are
pure real, with $Z_1(E) \rightarrow 1$ and $\phi_1(E) \rightarrow \Delta(T)$,
where $\Delta(T)$ is obtained self-consistently from the BCS equation. Clearly
then the quasiparticle energy is $E = \sqrt{\epsilon_k^2 + \Delta^2}$ and
the scattering rate, $\Gamma = 0$. Thus, within the BCS approximation in the
clean limit the
quasiparticle states are infinitely long-lived, as there is no means by 
which a quasiparticle can decay. With either impurity or phonon scattering,
quasiparticle decay becomes possible. In this way the BCS approximation in
the clean limit is pathological.\par

Let us first examine the situation where only electron-impurity scattering
is present, i.e. no phonon scattering occurs. Then the pairing and
renormalization function become
\def\sqq{\sqrt{\omega^2 Z^2(\omega) - \phi^2(\omega)}}
\def\sqqbcs{\sqrt{\omega^2  - \Delta^2}}
\begin{eqnarray}
\phi(\omega) & = & \phi_{cl}(\omega) + {i \over 2\tau} {\phi(\omega) \over\sqq}
\nonumber \\
Z(\omega) & = & Z_{cl}(\omega) + {i \over 2\tau} {Z(\omega) \over\sqq}
\label{impeqn}
\end{eqnarray}
\noindent where
the subscript `cl' refers to the clean limit and $1/\tau$ is the (normal)
impurity scattering rate. The gap function is defined
\begin{equation}
\Delta(\omega) = \phi(\omega)/Z(\omega)
\label{gapfunction}
\end{equation}
\noindent and is independent of impurity scattering. In the BCS limit,
\begin{eqnarray}
\phi(\omega) & = & \Delta + {i \over 2\tau} {\Delta \hbox{sgn} \omega 
\over\sqqbcs}
\nonumber \\
Z(\omega) & = & 1 + {i \over 2\tau} {\hbox{sgn} \omega \over\sqqbcs}.
\label{impeqnbcs}
\end{eqnarray}
\noindent As before, using the perturbative approach,
the quasiparticle energy is given by $E = \sqrt{\epsilon_k^2 + \Delta^2}$,
but the scattering rate is now (Pethick and Pines 1986)
\begin{equation}
\Gamma = {1 \over \tau} {\sqrt{E^2 - \Delta^2} \over | E | }
 = {1 \over \tau} {| \epsilon_k | \over \sqrt{\epsilon_k^2 + \Delta^2}}.
\label{gammaimpbcs}
\end{equation}
\noindent Eq. (\ref{gammaimpbcs}) shows that at the Fermi surface
impurities are ineffectual for electron scattering, i.e. this is another
manifestation of Anderson's theorem (Anderson 1959).
This happens abruptly at $T_c$; as soon as a gap develops the scattering 
rate becomes zero. Away from the Fermi surface the scattering rate is reduced
from its value in the normal state. In particular, well away from the Fermi
surface the scattering rate in the superconducting state approaches the normal
state rate. This behaviour is illustrated in Fig. 3.\par

The calculations in Fig. 3 rely on a perturbative search for the pole in
the lower half-plane, as given by Eq. (\ref{lifetime}). A more rigorous
calculation reveals poles ``within the gap'', but these do not appear
in the spectral function because of coherence factors. This can be seen by
rewriting Eq. (\ref{green}) as
\bea
G_{11}(k,\omega) & = & \phantom{+}{1 \over 2} \Biggl(1 + 
{\omega \over \sqqbcs}\Biggr) {1 \over
\sqqbcs - \epsilon_k + {i \over 2 \tau} \hbox{sgn} \omega }
\nonumber \\
& & - {1 \over 2} \Biggl(1 - {\omega \over \sqqbcs}\Biggr) {1 \over
\sqqbcs + \epsilon_k + {i \over 2 \tau} \hbox{sgn} \omega }
\label{greennew}
\eea
\noindent The coherence factors, $\Bigl(1 \pm {\omega \over \sqqbcs}\Bigr)$,
are such that the spectral function, $ A(k,\omega) \equiv - {1 \over \pi}
Im G(k,\omega)$, is zero for $| \omega | < \Delta$, independent of the impurity
scattering rate, as can be readily verified explicitly from 
Eq.~(\ref{greennew}). On the other hand the precise pole of the Green function
is given by solving for $\omega = E - i\Gamma/2$ in the two coupled
equations
\begin{equation}
E^2 - (\Gamma/2)^2 = \Delta^2 + \epsilon_k^2 - ({1 \over 2\tau})^2
\label{realpart}
\end{equation}
\begin{equation}
\Gamma = { | \epsilon_k | \over | E | } {1 \over \tau}
\label{imagpart}
\end{equation}
\noindent For $1/\tau = 0$ the solution is as before, $E = E_k \equiv
\sqrt{\epsilon_k^2 + \Delta^2}$, and $\Gamma = 0$. For finite impurity
scattering, however, the solution depends on ${1 \over 2\tau \Delta}$ at the 
Fermi surface ($\epsilon_k = 0$):
\begin{eqnarray}
E  =  \sqrt{\Delta^2 - ({1 \over 2\tau})^2}, \phantom{aaaaaaaaa}
\Gamma = 0 \phantom{aaaaaaaaa}\hbox{for} \phantom{aaaa}
{1 \over 2\tau \Delta} < 1
\label{ee1} \\
E  =  0,  \phantom{aaaaaaaaaaa}\Gamma = \sqrt{({1 \over \tau})^2 - (2\Delta)^2} 
\phantom{aaaa}\hbox{for}\phantom{aaaa}
{1 \over 2\tau \Delta} > 1
\label{ee2}
\end{eqnarray}
\noindent In either case the solution has a real part that lies
within the gap, that is, the quasiparticle residue at the pole is zero.
The ``relevant'' energies {\bff (i.e. for which the residue is non-zero)}
are $| E | > \Delta$, and then the scattering rate given by Eq. (\ref{imagpart})
agrees with the linearized solution, Eq. (\ref{lifetime}). {\bff However, the
quantity $\Gamma(E)$ is only physically meaningful when it is evaluated
at the pole energy, which occurs below the gap value, so in the ``relevant''
region $\Gamma(E)$ is merely a well-defined function.}
Away from the Fermi surface the quasiparticle energy requires the solution of
a quadratic equation and $E$ may lie below or above the gap, depending on 
$\epsilon_k$ and $1/\tau$. Once more solutions with $| E | < \Delta$ have
zero spectral weight. In Fig. 4a (b) we show the
real (imaginary) parts of the pole for some typical parameters. As can be
seen, some atypical behaviour can occur as a function of temperature,
for example when $\epsilon_k = 0$. As the temperature is lowered the pole
moves from a point on the negative imaginary axis to the origin 
(near $T/T_c = 0.8$) and then moves along the real axis towards 
$E = \sqrt{\Delta^2 - ({1 \over 2 \tau})^2}$. Nonetheless, comparison
of Fig. 4(b) with Fig. 3 shows that the perturbative is in qualitative agreement
with the non-perturbative solution.\par

Coherence factors are known to play an important role
for two-particle response functions (such as the NMR relaxation rate
or the microwave conductivity). In particular they lead to the
Hebel-Slichter (Hebel and Slichter 1959) singularity in the NMR relaxation rate.
However they have largely been ignored in single-particle functions.
Eq. (\ref{greennew}) shows, however, that the coherence factors play
a very important role in the single-particle spectral function, in
that they maintain a gap equal to $\Delta$, even when impurities
cause the single-particle pole to have a real part whose value
lies in the gap.\par

The frequency dependence of the
spectral function is shown in Fig. 5. For energies close to the Fermi surface 
the spectral function is dominated by the square root singularity at the gap 
edge, which comes from the coherence factors rather than the single-particle 
pole. For larger energies the peak present is due to the pole and differs
from the normal state spectral function only at low frequencies.\par

We return now to the strong coupling case. In the superconducting state,
we do not have access to the analytic continuation of the Green function
to the lower half-plane. (It is true
that we can attempt a representation through Pad\'e approximations,
as has often been done in the past, to analytically continue functions from
the imaginary axis to the real axis. However, these are notorious for
producing spurious poles (Marsiglio {\it et al.} 1988).)
We thus follow Karakozov {\it et al.} (1975),
and use an expansion near $\omega = 0$. We also follow
Kaplan {\it et al.} (1976) and henceforth use Eq. (\ref{lifetime})
for the scattering rate, noting that while the energy $E$ at which $\Gamma(E)$
is evaluated ought, in principle, to be computed self-consistently as was 
done in BCS, i.e. eqs. (\ref{realpart},\ref{imagpart}), in practice 
we will choose the relevant energy, and so treat energy as an independent 
variable.\par

Karakozov {\it et al.} (1975) pointed out that at any finite 
temperature the solutions to the Eliashberg equations have the following 
low frequency behaviour:
\begin{eqnarray}
Z(\omega) \approx  Z_1 + {i\gamma_2 \over \omega}
\label{zlow} \\
\Delta(\omega)  \approx  \delta_1 \omega^2 - i \delta_2 \omega
\label{gaplow}
\end{eqnarray}

\noindent where $Z_1, \gamma_2, \delta_1$ and $\delta_2$ are real
(positive) constants. This implies that a quasiparticle pole exists with
\begin{equation}
E \approx {\epsilon_k \over Z_1 \sqrt{1 + \delta_2^2}}
\label{enerlow}
\end{equation}

\begin{equation}
\Gamma \approx {1 \over Z_1} \Bigl( 4\pi \int_0^\infty d\nu \alpha^2F(\nu)
g(\nu) \bigl[ N(\nu) + f(\nu) \bigr] + {g(0) \over \tau } \Bigr)
\label{gammalow}
\end{equation}
\noindent where we have used Eq. (\ref{lifetime}) and assumed $E \rightarrow 0$
(i.e. we are near the Fermi surface). In Eq. (\ref{gammalow}) $g(\nu)$ is the
single electron density of states in the superconducting state,
\def\sqqbcsn{\sqrt{\nu^2  - \Delta^2}}
\begin{equation}
g(\nu) = Re \Biggl( {\nu \over \sqqbcsn} \Biggr),
\label{dos}
\end{equation}
\noindent which is non-zero for $\nu = 0$ at any finite temperature
(Karakozov {\it et al.} 1975; Marsiglio and Carbotte 1991; Allen and
Rainer 1991).
In Fig. 6 we show the quasiparticle
scattering rate, $\Gamma(T,E=0)$ vs. $T/T_c$ for various impurity
scattering rates, $1/\tau$, in the superconducting state. The normal 
state result is also shown for  reference. Note that in the clean limit
there is an enhancement just below $T_c$ in the superconducting state, but
at low temperatures there is an exponential suppression, compared to the
power law behaviour observed in the normal state. When impurity scattering 
is present there is an immediate suppression below $T_c$  (a knee is still
present, however, in the superconducting state). It is clear that at low 
temperatures the superconducting state is impervious to impurity scattering,
as one would expect. \par

Is $\Gamma(T,E=0)$ the relevant inverse lifetime by which properties
of the superconducting state can be understood ? Strictly speaking the
answer is no, particularly at low temperatures, where the spectral function
essentially develops a gap for low energies. In Fig. 7 we plot the spectral
function at
the Fermi surface, $A(k_F,\omega)$ for various temperatures
(Marsiglio and Carbotte 1991).
While no true gap exists at any finite temperature, it is clear that for low
temperatures the spectral weight at low frequency is exponentially suppressed.
Thus we have a situation similar to that in BCS theory, where
the {\bff quasiparticle pole occurs at an energy where the spectral weight
is essentially zero. It is more relevant to inquire about the scattering rate
(as defined by Eq. (\ref{lifetime})) evaluated
above the ``gap-edge'' where a large spectral weight is present, and
quasiparticles are more likely to be populated (as in detector applications).}
When impurities are added the spectral peak is broadened,
even at low temperatures, as shown in Fig. 8 for an intermediate
temperature, $T/T_c = 0.5$. However the lineshape is very asymmetric as a gap
remains at low frequencies (particularly prominent at low temperatures).\par

Fig. 7 demonstrates that the relevant energy is a function of temperature.
In Fig. 9 we plot $\Gamma(T,E=\Delta(T))$ vs. $T/T_c$, where $\Delta(T)$
is determined from the relation (at the Fermi surface)
\begin{equation}
E = Re \Delta(E,T).
\label{gap}
\end{equation}
\noindent By Eq. (\ref{gap}) we understand that the $E=0$ solution is excluded
(similarly the very low energy solution is also excluded). We are
interested in the conventional solution which gives rise to the peak in
the spectral function illustrated in Fig. 7 or 8. If no non-zero 
solution is present
(as is the case near $T_c$) then we utilize the $E=0$ solution. Note that
in this case care must be taken when obtaining the $E \rightarrow 0$ limit
of Eq. (\ref{lifetime}). Also there is present in this definition a
discontinuity at some temperature near $T_c$, which is where a nonzero
solution first appears. In this way
we hope to show the scattering rate at an energy where the spectral weight
is large, and therefore of most relevance to observables. At any rate, 
it is clear by 
comparing Fig. 9 to Fig. 6 (see also Fig. 2) that there is very little 
difference in the scattering rate in the gap region of energy. However, as the
energy increases beyond the gap the scattering rate increases, resulting
in short ``lifetimes'', even at zero temperature
(Kaplan {\it et al.} 1976). {\bff Thus,
while we have argued that the quasiparticle lifetime corresponding to the pole
of the Green function is of limited use because the quasiparticle residue there 
is zero, in practice the close correspondence between Fig. 9 and Fig. 6 
illustrates that it remains a useful indicator of the scattering rate for
energies up to about the gap energy.}

\section{THE OPTICAL CONDUCTIVITY}

\subsection{Extraction of a Scattering Rate from the Conductivity}

Using the Kubo formalism (Mahan 1981), the optical conductivity can be
related to a current-current correlation function. The final result for 
the frequency dependence of the conductivity in the long wavelength limit is
(Nam 1967; Lee {\it et al.} 1989; Bickers {\it et al.} 1990;
Marsiglio {\it et al.} 1992)
\begin{equation}
\sigma(\nu) = {i \over \nu} \Bigl( \Pi(\nu + i\delta) + {ne^2 \over m} \Bigr),
\label{cond}
\end{equation}
\noindent where the paramagnetic response function, $\Pi(\nu + i\delta)$, is
given in the previous article (Marsiglio and Carbotte 1997),
and will not be repeated here.
\par

To summarize the theoretical framework within which we are working:
for a given model for the spectral
density, $\alpha^2F(\nu)$ and a choice of impurity scattering rate $1/\tau$,
we can compute the conductivity $\sigma(\nu)$ at any frequency and temperature,
including effects due to both elastic and inelastic scattering mechanisms, the
latter being determined by the choice of electron-phonon spectral density. 
As explained in the introduction, however, vertex corrections are
omitted.\par

In the analysis of experimental data it is possible to use several methods
to extract a single temperature dependent scattering rate. One such method
utilized the microwave conductivity in \YBCO (Bonn {\it et al.} 1992, 1993)
and in Nb (Klein 1994), and adopted a two-fluid model description of the 
superconducting state.
It was assumed that the absorptive component of the conductivity (real part of
$\sigma$ at finite frequency, denoted by $\sigma_1(\nu)$) was due only to the
normal component of the fluid. Thus an expression of the form
\be
\sigma_1(\nu,T) = {ne^2 \over m} {m \over m^\ast(T)} 
\Biggl[ 1- {\lambda^2(0) \over \lambda^2(T)}
\Biggr] {\tau(T) \over 1 + (\nu\tau(T))^2}
\label{twofluid}
\ee
\noindent is assumed to hold approximately. In Eq. (\ref{twofluid}) $\tau(T)$
has units of a scattering time and $\lambda(T)$ is the penetration depth
at temperature $T$. Here we will calculate $\sigma_1(\nu,T)$ for a model 
$\alpha^2F(\nu)$ and $1/\tau$ using the full expression (\ref{cond}). At
the same time we can calculate the penetration depth, either from a zero
frequency limit of Eq. (\ref{cond}), or directly from the imaginary axis
(Nam 1967; Marsiglio {\it et al.} 1990):
\be
{\lambda^2(0) \over \lambda^2(T)} = \pi T \sum_{m=-\infty}^\infty
{\phi_m^2 \over (\omega^2_m Z^2(i\omega_m) + \phi_m^2)^{3/2} }.
\label{penet}
\ee
\noindent The idea is to examine the zero frequency limit of 
Eq. (\ref{twofluid}), and thus define a scattering rate relative to the
rate at $T_c$ (Klein 1994):
\be
{\tau(T_c) \over \tau(T)} \equiv \Biggl(1 - {\lambda^2(0) \over \lambda^2(T)}
\Biggr)  {\sigma_N(T_c) \over \sigma_1(T)},
\label{klein_onet}
\ee
\noindent where it has been assumed that the mass enhancement factor in
Eq. (\ref{twofluid}) does not change with temperature. {\bff In this way
an effective scattering rate was extracted from microwave data for
\YBCO
(Bonn {\it et al.} 1992, 1993)
and Nb (Klein 1994).}\par

{\bff To see how closely this scattering rate follows the quasiparticle
scattering rate, we show results} in Fig. 10 for $\tau(T_c)/\tau(T)$ defined by 
Eq. (\ref{klein_onet})
in the superconducting state (solid curve) with which we compare the
inverse quasiparticle lifetime for both the normal (short-dashed curve, 
Eq. (\ref{gammak}) ),
and the superconducting (long-dashed curve, Eq. (\ref{lifetime}) ) states.
The results are
based on a Debye model spectrum for $\alpha^2F(\nu)$ used in the previous 
section, with mass enhancement parameter $\lambda = 1$ and $T_c = 100$ K. (A
negative $\mu^\ast$ is required.) Note that the normal state scattering rate
(short-dashed curve) is only sensitive to the low frequency part of
$\alpha^2F(\nu)$ at low temperatures: a $\nu^2$ dependence in $\alpha^2F(\nu)$
implies a $T^3$ dependence in $1/\tau(T)$ (Again, vertex corrections
would alter this to the familiar $T^5$ law (Ashcroft and Mermin 1976).)
The agreement between the inverse 
lifetime (long-dashed curve) and the scattering rate defined by the 
two-fluid model (solid curve) in the
superconducting state is remarkable, although later we shall see that this
quantitative agreement occurs only with this particular Debye spectrum.
This indicates that the two-fluid
description makes sense (Berlinsky {\it et al.} 1993; Bonn {\it et al.} 1994)
at least qualitatively, and the quantities shown in 
Fig. 10 apply to the normal component of the superfluid. 
Fig. 10 illustrates the comparison in the clean limit, where the
two-fluid description is expected to be most accurate (Bonn {\it et al.} 1994).
Before
investigating impurity dependence, we turn to a second possible procedure
for extracting a scattering rate from conductivity data
(Romero {\it et al.} 1992; Tanner and Timusk 1992),
which is simply a generalization of that used by Shulga {\it et al.} (1991)
to the superconducting state. One simply fits the low frequency
absorptive part of the conductivity to a Drude form:
\be
\sigma_1(\nu,T) = {ne^2 \over m}{1 \over m^\ast/m} {\tau^\ast(T) \over
1 + (\nu\tau^\ast)^2}.
\label{romerofit}
\ee
\noindent As described in 
Shulga {\it et al.} (1991) and Marsiglio and Carbotte (1995),
it is possible to fit a Drude form to the low frequency 
part of the optical conductivity in the normal state. Such a fit is also
possible in the superconducting state (Romero {\it et al.} 1992).
Theoretically, the fit
is problematic in a BCS approach because there is a Hebel-Slichter
logarithmic singularity at low frequency {\it at all temperatures in the
superconducting state}. However, with the Eliashberg approach, the
Hebel-Slichter singularity is smeared, and one can fit a Drude form over
a limited range of frequency. Such a fit for a Debye spectrum is also included
in Fig. 10 (dotted curve). While the fit is in qualitative agreement
with the inverse lifetime, this method of characterizing the inverse lifetime
in the superconducting state is clearly not as accurate as the two-fluid
model. The fits themselves are shown in Fig. 11. It is
clear that the fits fail at sufficiently high frequency, as one would expect,
but that they characterize well the low frequency response in the 
superconducting state. \par

Fig. 10 clearly shows that there is considerable freedom and hence ambiguity
in extracting a temperature dependent scattering rate from conductivity data.
Nevertheless, in the clean limit it is evident that the two-fluid model
is a useful device for extracting the low energy inverse quasiparticle lifetime 
in the superconducting state.\par

With the addition of impurities the situation changes considerably. This is 
illustrated in Fig. 12, where the same calculations as in Fig. 10 are
shown, but with an additional impurity scattering, $1/\tau = 25$ meV,
included. The use of formula (\ref{twofluid}), inspired by the 
two-fluid model,
gives a scattering rate (solid curve) that falls much less rapidly
around $T=T_c$ then does the inverse quasiparticle
lifetime in the superconducting state
(long-dashed curve). The latter curve drops almost vertically as the 
temperature drops below $T_c$, as has already been discussed.
The result based on the Drude fit (dotted curve) shows a peak
which is reduced in size from that in the clean limit (Fig. 10). Indeed,
for increased impurity scattering, the peak just below $T_c$ disappears.
In both cases a rapid suppression is expected just below $T_c$: in the 
case of the inverse lifetime, this suppression is a consequence of Anderson's
theorem (Anderson 1959), as already discussed. In the case of the Drude fit,
it is easy to see that this is the case in the dirty limit. In the dirty limit
the conductivity  is almost flat as a function of frequency on the scale
of $1/\tau$, at $T_c$. Just below $T_c$, however, a gap in the spectrum
begins to develop, so that weight is shifted from roughly the gap region
to the delta function at the origin. Any low energy fit will then use
a Lorentzian width which senses this depression in the conductivity, which is
on an energy scale of the gap. This represents a significant suppression
from the normal state scattering rate (infinite in the dirty limit). Here
we are in an intermediate regime, with $1/\tau = 25$ meV (note: $\Delta(T=0)
= 20.2 $ meV). The corresponding fits for Fig. 12 are shown in Fig. 13.

\subsection{Possible Origin of the Low Temperature Conductivity Peak}

As an aside we wish to further
examine the low frequency conductivity vs. reduced temperature,
a quantity measured in the high $T_c$ oxides by microwave techniques
(Nuss {\it et al.} 1991; Bonn {\it et al.} 1992, 1994).
{\bff In these experiments a peak was observed
in the real part of the low frequency conductivity as a function of temperature,
the origin of which has been discussed in many contexts. Here, we add yet
another possibility}. In
Fig. 14 we show the real part of the conductivity, $\sigma_1(\nu)$ vs.
reduced temperature, for several low frequencies. {\bff We continue to
use a model Debye spectrum to simply establish qualitative effects.}
Note the relative 
insensitivity to frequency, a feature of strong coupling pointed out
in Marsiglio (1991). Also note the lack of a coherence peak just
below $T_c$. Nonetheless, a broad peak exists at lower temperatures,
somewhat reminiscent of that observed in \YBCO
(Nuss {\it et al.} 1991; Bonn {\it et al.} 1992, 1994).
This peak exists because of a competition between an increasing scattering time
(making $\sigma$ increase) and a decreasing normal component (making $\sigma$
decrease, particularly as $T \rightarrow 0$). We should note that this peak
is most prominent in the clean limit. As Fig. 13 indicates (see values at 
the intercept), the peak is absent for a sufficiently large impurity scattering
rate. \par

It is of interest to examine what dependence these results have on
the electron-phonon spectral function. As an extreme we utilize
a spectrum which is sharply peaked at some high frequency, and, in contrast
to the Debye model employed above,
coupling to low frequency modes is absent; such
a spectrum models a strong coupling to an optic mode. We choose a triangular
shape for convenience, starting at $\omega_0 = 34.8$ meV with a cutoff at
$\omega_E = 35.5$ meV. The coupling constant is chosen so that the
mass enhancement value is $\lambda = 1$, in agreement with that chosen for
the Debye spectrum. As was the case there, a negative $\mu^\ast$ is used to
give $T_c = 100$ K, and the zero temperature gap was found to be close
to the Debye value.\par

In Fig. 15 we plot the real part of the low frequency conductivity, 
$\sigma_1(\nu)$ vs. reduced temperature, now using the  triangular spectrum
for $\alpha^2F(\nu)$. In contrast to the result for the Debye model, the low
frequency conductivity is strongly frequency dependent at low temperatures.
In fact, for $\nu = 0$, the conductivity appears to diverge.  We believe that
at sufficiently low temperature, this curve will actually achieve a maximum
and approach zero at zero temperature, but we have been unable to obtain this
result numerically. Once again there is a competition between an increasing
scattering time and a decreasing normal density component as the 
temperature is lowered from $T_c$. Here, however, the increasing scattering
time appears to be overwhelming the decrease in normal fluid density. The
key difference with the Debye spectrum is that here the spectrum has a big gap, 
so that the lifetime is increasing exponentially with decreasing temperature,
already in the normal state. Recall that in the Debye case the increase
followed a power law behaviour with decreasing temperature. Since the decrease
in normal fluid density is always exponential, this term dominates
in the Debye case,
whereas in the case of the gapped spectrum the competition is subtle, and 
will depend strongly on the details of the spectrum (an electron-phonon
spectrum with a much smaller gap will yield a zero frequency conductivity
which approaches zero at zero temperature, for example).\par

The physics of this conductivity peak is different from what has been already
proposed for \YBCO. It has been suggested that the excitation spectrum
{\em becomes gapped due to the superconductivity}, i.e. a feedback
mechanism exists which creates a low frequency gap in the excitation 
spectrum as the superconducting
order parameter opens up below $T_c$. Such a scenario has been explored within
a marginal Fermi liquid scheme
(Nuss {\it et al.} 1991; Nicol and Carbotte 1991a, 1991b), and is also
consistent with thermal conductivity experiments (Yu {\it et al.} 1992).
Here the 
conductivity peak arises because the spectrum is already gapped, and
the scattering rate is sufficiently high at $T_c$ because $T_c$ itself
is fairly high. We should warn the reader that this mechanism requires
a ``fine tuning'' of the spectrum, i.e. a large gap is required {\bff in
the phonon spectrum. An alternate boson spectrum may be operative, and then a
less extreme boson spectrum may work, if the normal state
electron density falls at a correspondingly slower rate as the temperature is
lowered. This would be achieved if the superconducting order parameter
has nodes, for example (Schachinger and Carbotte 1996).}\par

Note that for any non-zero frequency the conductivity has a visible maximum, 
and quickly approaches zero at sufficiently low temperature. Nonetheless
this turn around occurs for yet another reason: as one
lowers the temperature the Drude-like peak at low frequencies gets
narrower while at the same time the magnitude of the zero temperature
intercept increases. For any given finite frequency, then, a temperature
is eventually reached below which this frequency is now on the tail
of the Drude-like peak. This means that
while the zero frequency conductivity
increases, that at any finite frequency will eventually decrease
as the width becomes smaller than the frequency.
So in this case the increase in scattering
lifetime still dominates the decrease in normal fluid density, but,
because we are fixed at a finite frequency, the conductivity
decreases. Note that frequencies of order 0.01 meV
($\approx 2.4$ GHz) are within the range of microwave frequencies that are
used in experiments.\par

To show this more explicitly  we illustrate in Fig. 16 the various 
scattering rates obtained with the gapped spectrum. These are to be compared
with those shown for the Debye model in Fig. 10. Clearly the quantitative
agreement between the scattering rate inspired by the two-fluid model 
and the
inverse lifetime that was seen in Fig. 10 with the Debye spectrum was
fortuitous. While a qualitative correspondence between these two entities
continues to exist with the gapped spectrum, they are no longer in 
quantitative agreement. We have verified that this is generically
true, by investigating other spectra, not shown here.

\section{SUMMARY}

We have investigated the quasiparticle lifetime in an Eliashberg s-wave
superconductor, generalizing earlier work (Kaplan {\it et al.} 1976) to include
impurity scattering as well. We find that the quasiparticle lifetime
becomes infinite at low temperatures, independent of the impurity
scattering rate, which we understand as simply a manifestation of
Anderson's theorem (Anderson 1959). Thus, on general grounds, within
a BCS framework, the inverse quasiparticle lifetime should collapse to zero
in the superconducting state. {\bff We have also emphasized the importance of
the so-called coherence factors for the single-particle spectral function,
when impurity scattering is included within the BCS case.}\par

Two methods have been investigated for extracting the scattering rate from
the low frequency conductivity. One relies on a two-fluid model
picture, and the other simply utilizes a low frequency Drude fit. Neither
should necessarily correspond very closely to the quasiparticle
inverse lifetime, and we find that in general they do not, quantitatively.
Qualitatively, however, the scattering rate defined by either procedure
gives the correct temperature dependence for the inverse lifetime. In the 
presence of impurities, the two-fluid prescription appears to be less accurate,
presumably because such a prescription takes into account only the lower
normal fluid density as the temperature is lowered, and not the fact
that impurity scattering is less effective in the superconducting state.
Thus we caution that the interpretation of a conductivity-derived
scattering rate as a quasiparticle inverse lifetime, while qualitatively
correct, is quantitatively inaccurate.
\par
%\begin{thebibliography} {999}
\section*{REFERENCES}

\vskip0.5cm
\noindent
$^\dagger$ Present address: Dept. of Physics, University of Alberta,
Emonton, Canada T6G 2J1 \par

\vskip0.5cm
\noindent
Allen, P.B., and Mitrovi\'{c}, B. (1982). In {\it Solid State Physics,}
edited by H. Ehrenreich, F.~Seitz, and D. Turnbull (Academic, New York)
Vol. 37, p.1.\par

\vskip0.5cm
\noindent
Allen, P.B. and Rainer, D. (1991). Nature {\bf 349}, 396.\par

\vskip0.5cm
\noindent
Anderson, P.W. (1959). J. Phys. Chem. Solids {\bf 11}, 26.\par

\vskip0.5cm
\noindent
Ashcroft, N.W., and Mermin, N.D. (1976) {\it Solid State Physics}
(Saunders College, Holt, Rinehart and Winston, Philadelphia).\par

\vskip0.5cm
\noindent
Bedell, K.S., Coffey, D., Meltzer, D.E.,
Pines, D. and Schrieffer, J.R. (1990). See the various reviews in
{\it High Temperature Superconductivity:
The Los Alamos Symposium}, (Addison-Wesley, Don Mills).\par

\vskip0.5cm
\noindent
Berlinsky, A.J., Kallin, C., Rose, G. and Shi, A.-C. (1993).
Phys. Rev. B {\bf 48}, 4074. There is a typographical error
in Eqs. (15) and (16).\par

\vskip0.5cm
\noindent
Bickers, N.E., Scalapino, D.J., Collins, T., and Schlesinger, Z. (1990).
Phys. Rev. B{\bf 42}, 67.\par

\vskip0.5cm
\noindent
Bonn, D.A., Dosanjh, P., Liang, R. and Hardy, W.N. (1992). Phys. Rev. Lett.
{\bf 68}, 2390.\par

\vskip0.5cm
\noindent
Bonn, D.A., Liang, R., Riseman, T. M., Baar, D.J., Morgan, D.C.,
Zhang, K., Dosanjh, P., Duty, T.L., MacFarlane, A., Morris, G.D., Brewer, J.H.,
Hardy, W.N., Kallin, C. and Berlinsky, A.J. (1993). Phys. Rev. B{\bf 47},
11314.\par

\vskip0.5cm
\noindent
Bonn, D.A., Zhang, K., Kamal, S., Liang, R., Dosanjh, P., Hardy, W.N.,
Kallin, C. and Berlinsky, A.J. (1994). Phys. Rev. Lett. {\bf 72}, 1391.\par

\vskip0.5cm
\noindent
Eliashberg, G.M. (1960). Zh. Eksperim. i Teor. Fiz. {\bf 38}, 966; Soviet Phys.
JETP {\bf 11}, 696.\par

\vskip0.5cm
\noindent
Grimvall, G. (1981). {\it The Electron-Phonon Interaction in Metals}
(North-Holland, New York).\par

\vskip0.5cm
\noindent
Hebel, L.C. and Slichter, C.P. (1959). Phys. Rev. {\bf 113}, 1504.\par

\vskip0.5cm
\noindent
Kaplan, S.B., Chi, C.C., Langenberg, D.N., Chang, J.J., 
Jafarey, S. and Scalapino, D.J. (1976). Phys. Rev. B{\bf 14}, 4854.\par

\vskip0.5cm
\noindent
Karakozov, A.E., Maksimov, E.G. and Mashkov, S.A. (1975).
Zh. Eksp. Teor. Fiz. {\bf 68},
1937; Sov. Phys. JETP {\bf 41}, 971 (1976).\par

\vskip0.5cm
\noindent
Klein, O. (1994). Phys. Rev. Lett. {\bf 72}, 1390.\par

\vskip0.5cm
\noindent
Lee, W., Rainer, D., and Zimmermann, W. (1989). Physica C{\bf 159}, 535.\par

\vskip0.5cm
\noindent
Mahan, G.D. (1981). {\it Many-Particle Physics} (Plenum Press, New York);
(1987) Phys. Reports {\bf 145}, 251.\par

\vskip0.5cm
\noindent
Marsiglio, F., Schossmann, M. and Carbotte, J.P. (1988). Phys. Rev. B{\bf 37}, 
4965.\par

\vskip0.5cm
\noindent
Marsiglio, F. Carbotte, J.P. and Blezius, J. (1990). Phys. Rev. B{\bf 41},
6457; J.P. Carbotte, Rev. Mod. Phys. {\bf 62}, 1027 (1990).\par

\vskip0.5cm
\noindent
Marsiglio, F. and Carbotte, J.P. (1991). Phys. Rev. B{\bf 43}, 5355.\par

\vskip0.5cm
\noindent
Marsiglio, F. (1991). Phys. Rev. B{\bf 44}, 5373.\par

\vskip0.5cm
\noindent
Marsiglio, F. Akis, R., and Carbotte, J.P. (1992). Phys. Rev. B{\bf 45}, 9865.
The derivation for
the phonon self-energy at $q = 0$, applies to the optical conductivity.\par

\vskip0.5cm
\noindent
Marsiglio, F. and Carbotte, J.P. (1995). Phys. Rev. B{\bf 52}, 16192.\par

\vskip0.5cm
\noindent
Marsiglio, F., Carbotte, J.P., Puchkov, A. and Timusk, T. (1996). Phys. Rev. B
{\bf 53}, 9433.\par

\vskip0.5cm
\noindent
Marsiglio, F. and Carbotte, J.P. (1997) previous article.\par

\vskip0.5cm
\noindent
Nam, S.B. (1967). Phys. Rev. B{\bf 156}, 470; Phys. Rev. B{\bf 156}, 487.\par

\vskip0.5cm
\noindent
Narayanamurti, V., Dynes, R.C., Hu, P., Smith, H. and Brinkman, W.F. (1978).
Phys. Rev. B{\bf 18}, 6041.\par

\vskip0.5cm
\noindent
Nicol, E.J. (1991). Ph.D. thesis, McMaster University, unpublished.\par

\vskip0.5cm
\noindent
Nicol, E.J. and Carbotte, J.P. (1991a). Phys. Rev. B{\bf 44}, 7741.\par

\vskip0.5cm
\noindent
Nicol, E.J. and Carbotte, J.P. (1991b). Physica C{\bf 185}, 162.\par

\vskip0.5cm
\noindent
Nuss, M.C., Mankiewich, P.M., O'Malley, M.L., Westerwick, E.H. and
Littlewood, P.B. (1991). Phys. Rev. Lett. {\bf 66}, 3305.\par

\vskip0.5cm
\noindent
Owen, C.S. and Scalapino, D.J. (1971). Physica (Amsterdam), {\bf 55}, 691.\par

\vskip0.5cm
\noindent
Pethick, C.J. and Pines, D. (1986). Phys. Rev. Lett. {\bf 57}, 118.\par

\vskip0.5cm
\noindent
Rainer, D. and Bergmann, G. (1974). J. Low Temp. Phys. {\bf 14}, 501.\par

\vskip0.5cm
\noindent
Romero, D.B., Porter, C.D., Tanner, D.B., Forro, L., Mandrus, D., Mihaly, L.,
Carr, G.L. and Williams, G.P. (1992). Phys. Rev. Lett. {\bf 68}, 1590.\par

\vskip0.5cm
\noindent
Scalapino, D.J. (1969). In {\it Superconductivity},
edited by R.D. Parks (Marcel Dekker, New York) Vol. 1, p.449.\par

\vskip0.5cm
\noindent
Schachinger, E. and Carbotte, J.P. (1996). Proceedings of the MOS conference
(Karlsruhe, 1996)in press.

\vskip0.5cm
\noindent
Schrieffer, J.R. (1983). {\it Theory of Superconductivity} (Benjamin/Cummings,
Don Mills), p.122.\par

\vskip0.5cm
\noindent
Shulga, S.V., Dolgov, O.V., and Maksimov, E.G. (1991). Physica C
{\bf 178}, 266; Dolgov, O.V., Maksimov, E.G.,  and Shulga, S.V. (1991).
In {\it Electron-Phonon Interaction in Oxide Superconductors}, edited by
R. Baquero (World Scientific,
Singapore), p. 30.\par

\vskip0.5cm
\noindent
Tanner, D.B., and Timusk, T. (1992). In {\it Physical Properties of
High Temperature Superconductors III}, edited by D.M. Ginsberg
(World Scientific, Singapore) p. 363.\par

\vskip0.5cm
\noindent
Yu, R.C., Salamon, M.B., Lu, J.P. and Yee, W.C. (1992). Phys. Rev. Lett.
{\bf 69}, 1431.

%\end{thebibliography}

\vfil\eject
%\bigskip
%\bigskip
%\noindent {\bf Figure Captions}
%\par
\begin{figure}
\psfig{file=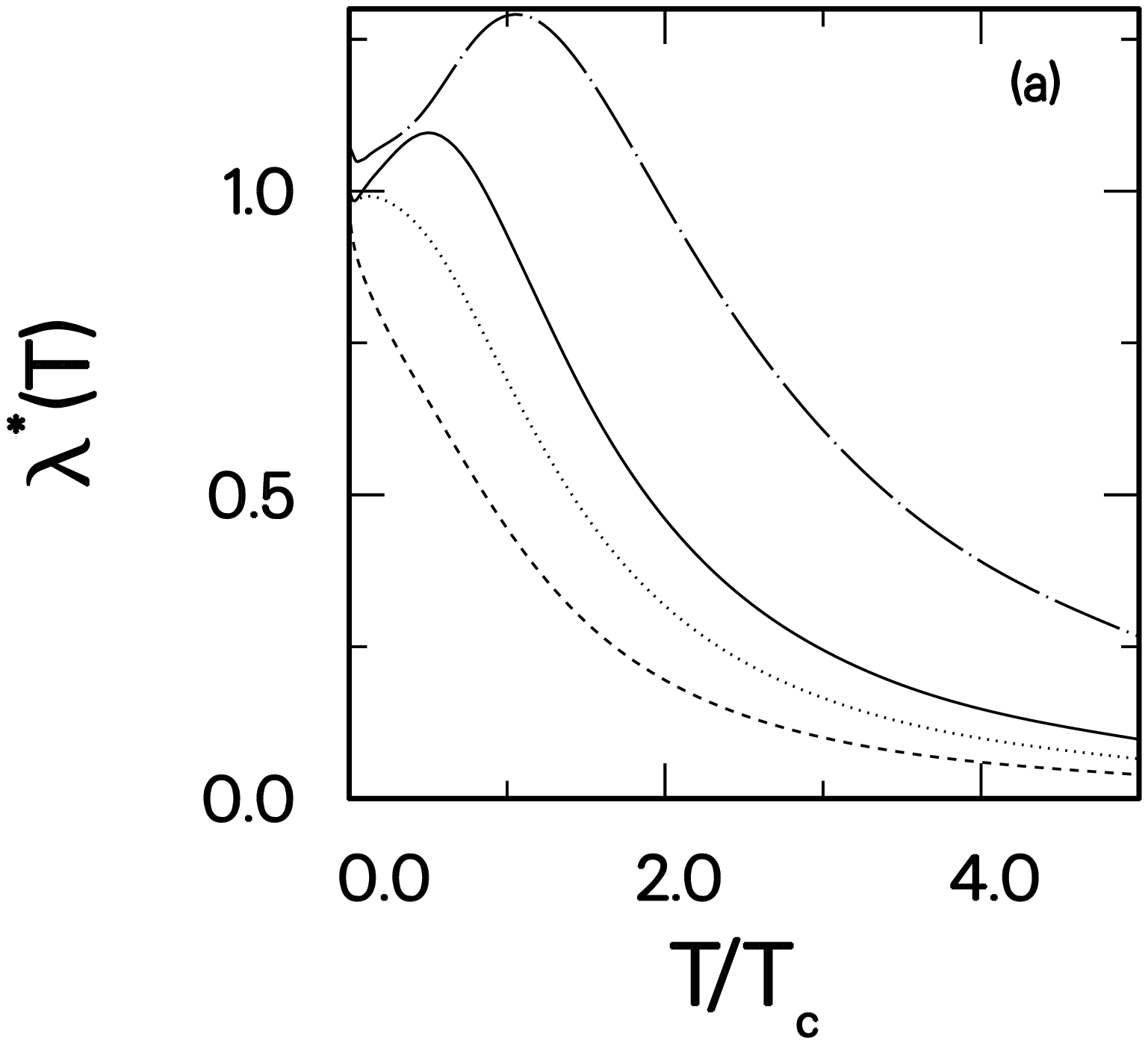,height=7.5in,width=6.7in}
\vskip+0mm
\caption{(a) Mass enhancement parameter, as defined by 
Eq. (\protect\ref{lambdastar}) in the text, vs. reduced 
temperature $T/T_c$, for various
electron-phonon spectral functions. In all cases $T_c = 100$ K. The 4
spectra used are a Debye spectrum (solid line), linear spectrum (dotted line),
a spectrum proportional to $\protect\sqrt{\nu}$ (dashed line), and a triangular
spectrum (dot-dashed line). In all cases the strength is such that 
$\protect\lambda = 1$. In the first three cases a cutoff frequency
equal to 30 meV
was used. In the last case, the spectrum starts at 34.8 meV and is cut off
at 35.5 meV.
(b) Inverse lifetime, $\protect\Gamma(T)$ (in meV) vs. reduced temperature for
the same spectra as in (a). Note the different low temperature behaviour, 
depending on the low frequency characteristics of the electron-phonon
spectrum used.}
\end{figure}
\vfil\eject

\begin{figure}
\psfig{file=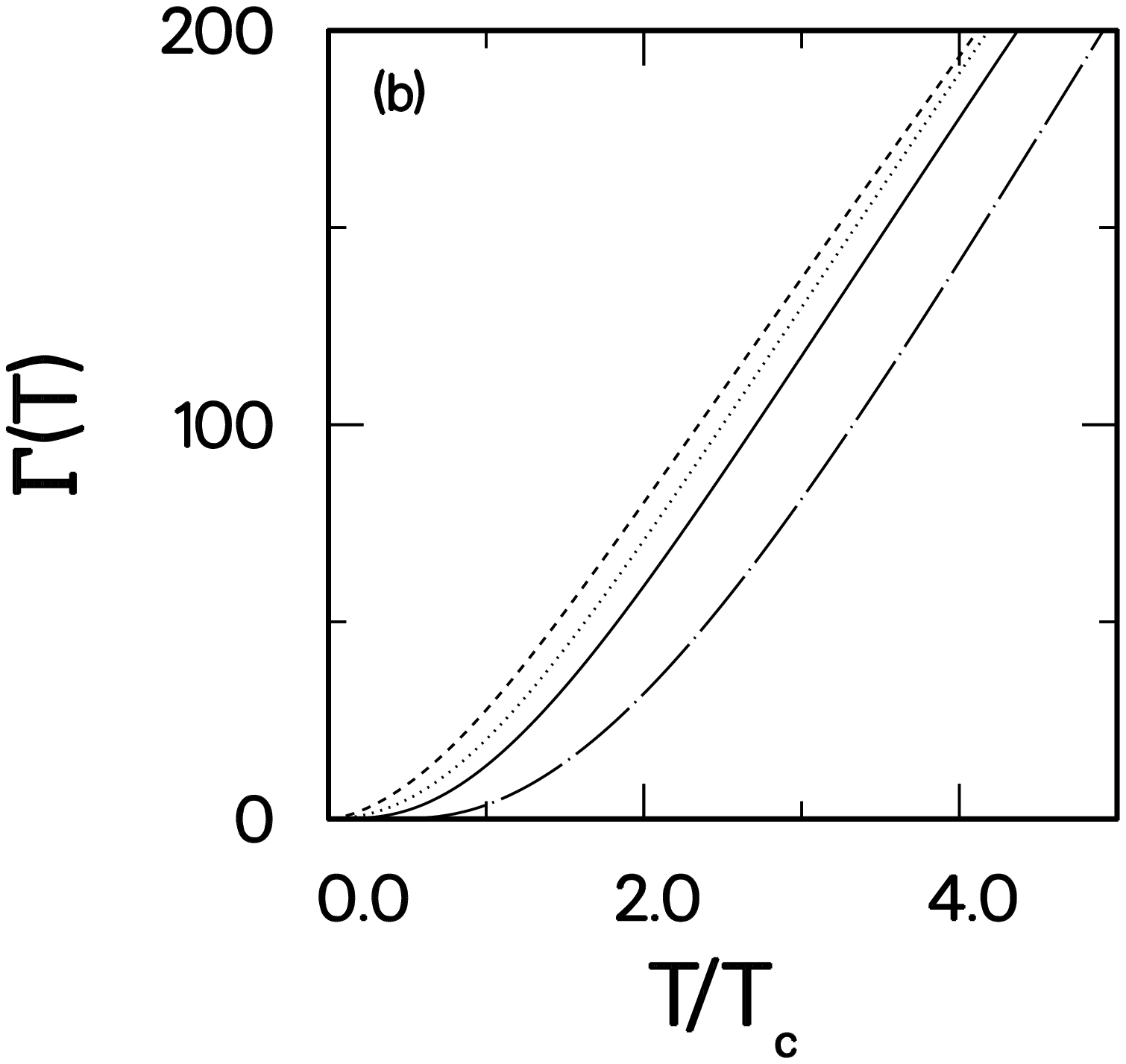,height=7.5in,width=6.7in}
\vskip+0mm
\end{figure}
\vfil\eject

\begin{figure}
\psfig{file=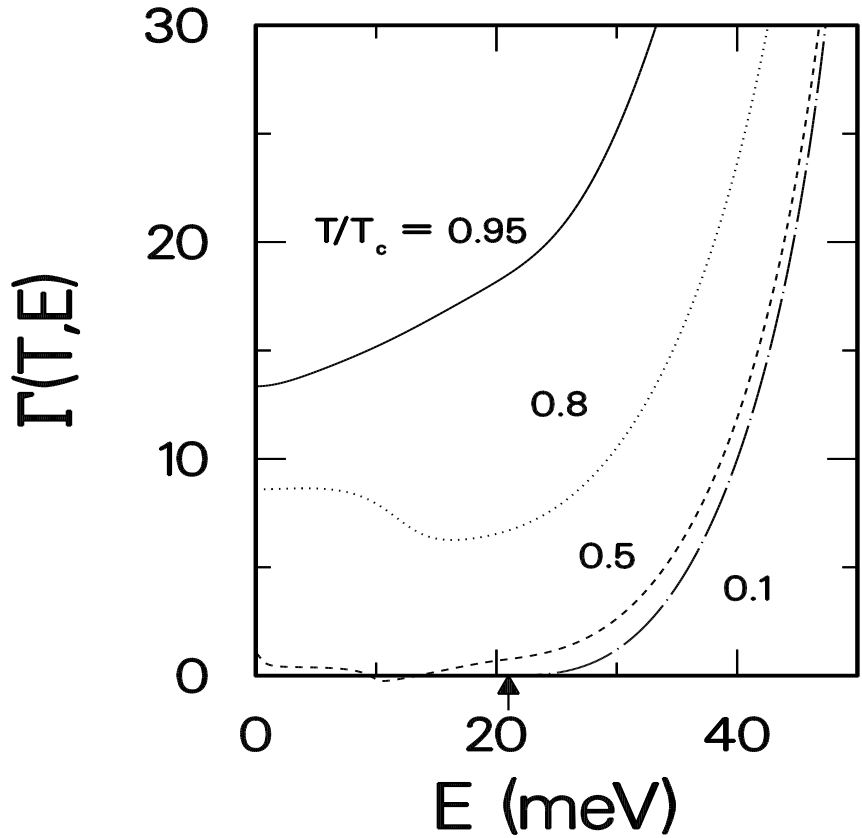,height=7.5in,width=6.7in}
\vskip+0mm
\caption{Scattering rate, $\Gamma(T,E)$ (in meV) vs. E (in meV) for
various temperatures in the superconducting state. The zero temperature
gap at the Fermi surface is indicated by the arrow. Note that for $T/T_c= 0.5$
the function plotted actually becomes negative (near 10 meV). The physical
pole occurs at an energy, E,  where $\Gamma(T,E)$ is always positive, however.}
\end{figure}
\vfil\eject

\begin{figure}
\psfig{file=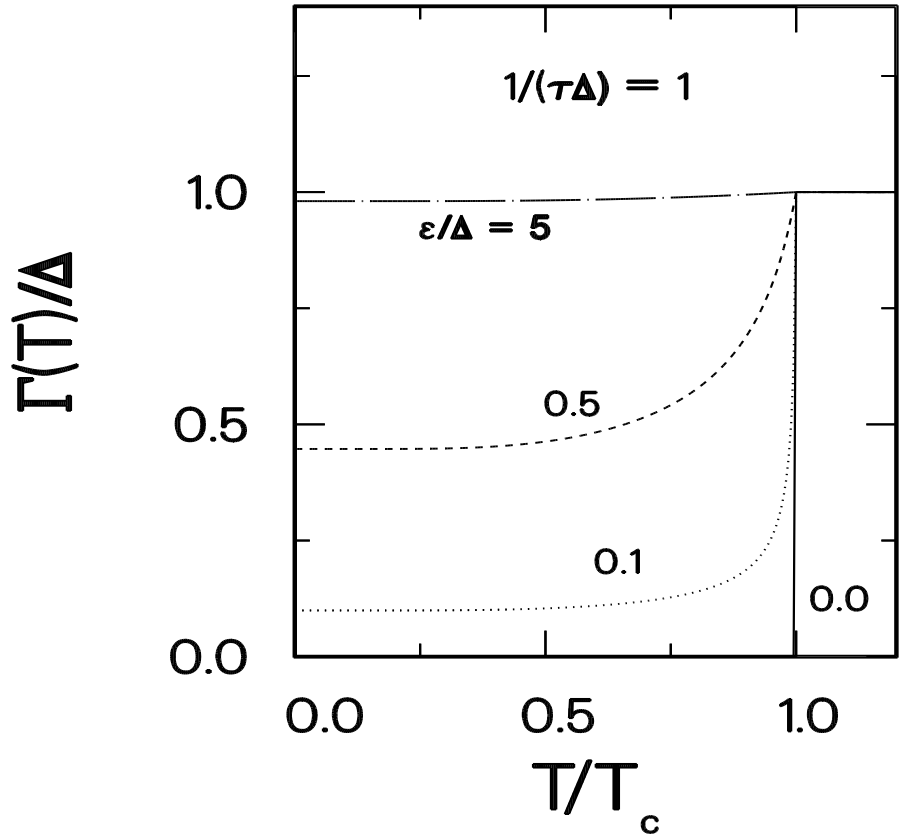,height=7.5in,width=6.7in}
\vskip+0mm
\caption{Scattering rate normalized to the zero temperature gap,
$\Gamma(T,E)/\Delta$ vs. reduced temperature $T/T_c$, for various
quasiparticle energies, $\epsilon$. These are computed using the
perturbative expansion, Eq. (\protect\ref{gammaimpbcs}). 
The impurity scattering rate is
$1/(\tau \Delta) = 1$. Note that on the Fermi surface (solid curve), the
scattering rate is zero immediately below $T_c$.}
\end{figure}
\vfil\eject
\begin{figure}
\psfig{file=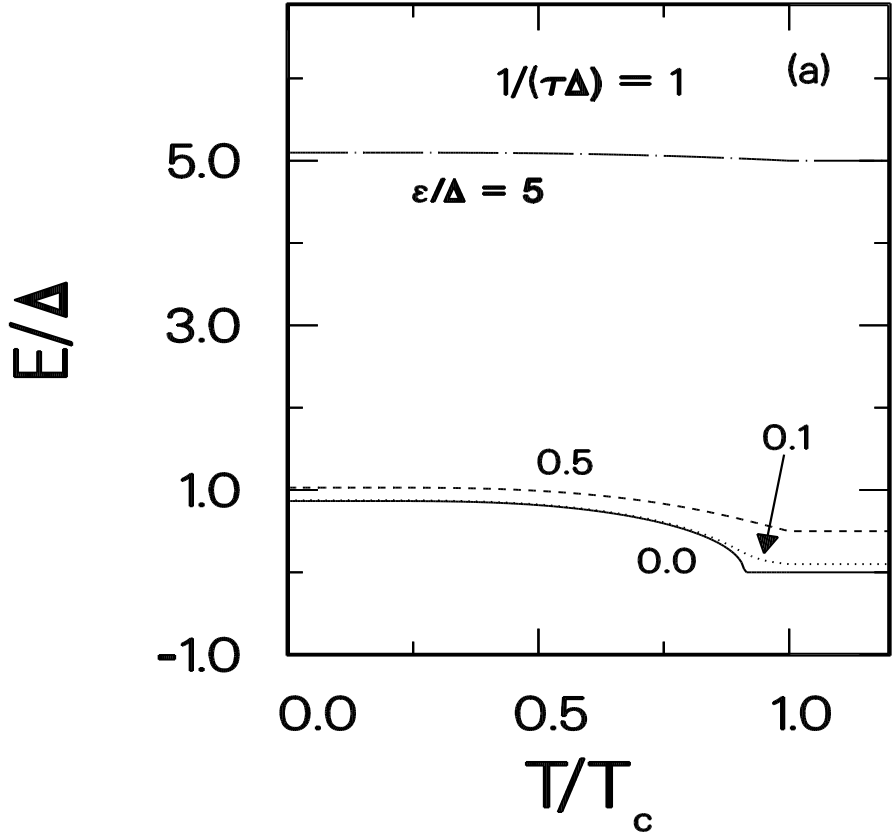,height=7.5in,width=6.7in}
\vskip+0mm
\caption{(a) Real and (b) imaginary parts of the quasiparticle pole
vs. reduced temperature, $T/T_c$, with $1/(\tau \Delta) = 1$ and various
quasiparticle energies. These are computed non-perturbatively from
Eqs. (\protect\ref{realpart},\protect\ref{imagpart}).
Note that  at the Fermi surface the 
quasiparticle energy has an abrupt onset at a temperature somewhat 
{\em below} $T_c$. The results in (b) are similar to those in Fig. 3, except
that at the Fermi surface, the decrease in scattering rate below $T_c$
is not as abrupt.}
\end{figure}
\vfil\eject
\begin{figure}
\psfig{file=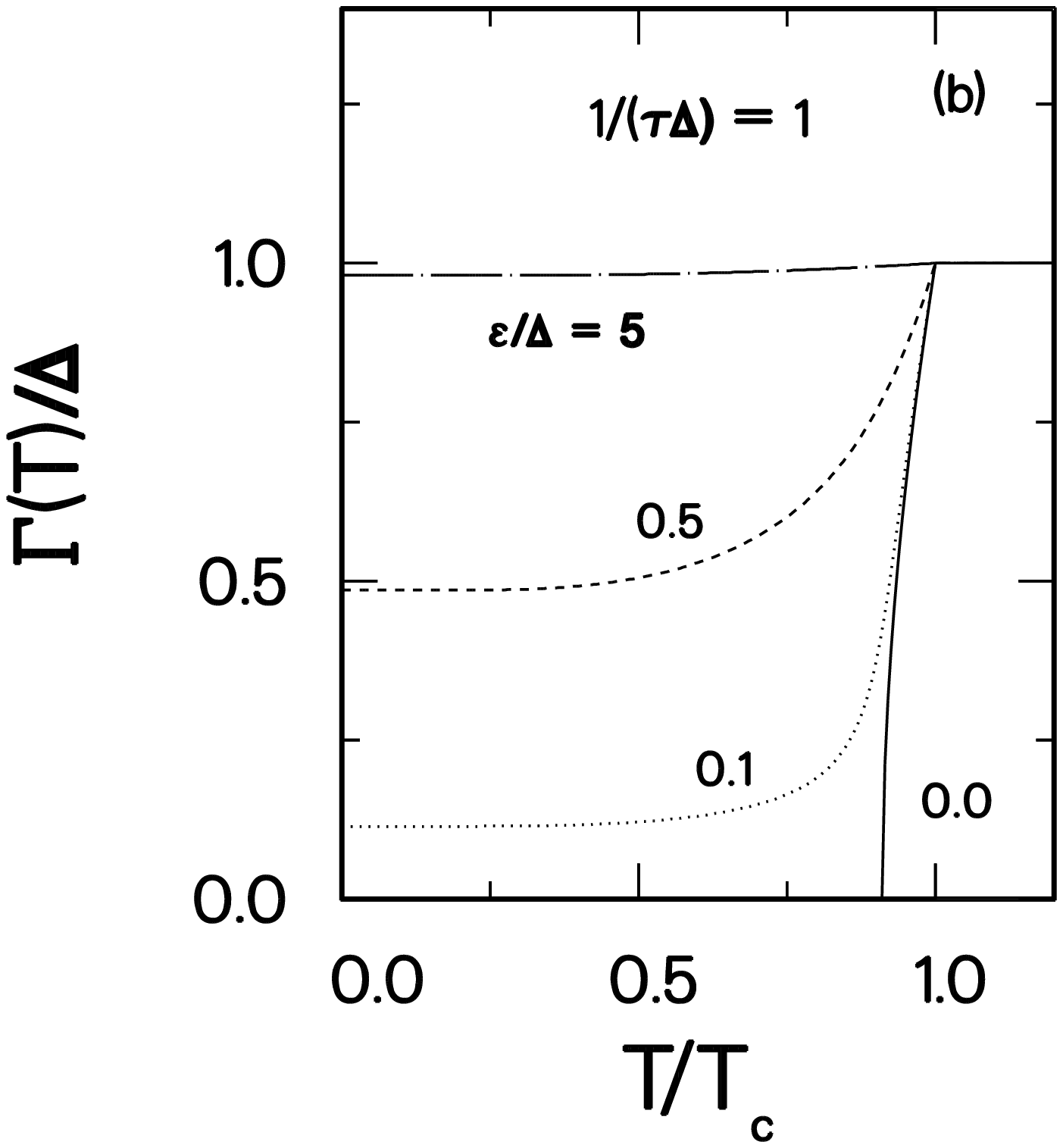,height=7.5in,width=6.7in}
\vskip+0mm
\end{figure}
\vfil\eject

\begin{figure}
\psfig{file=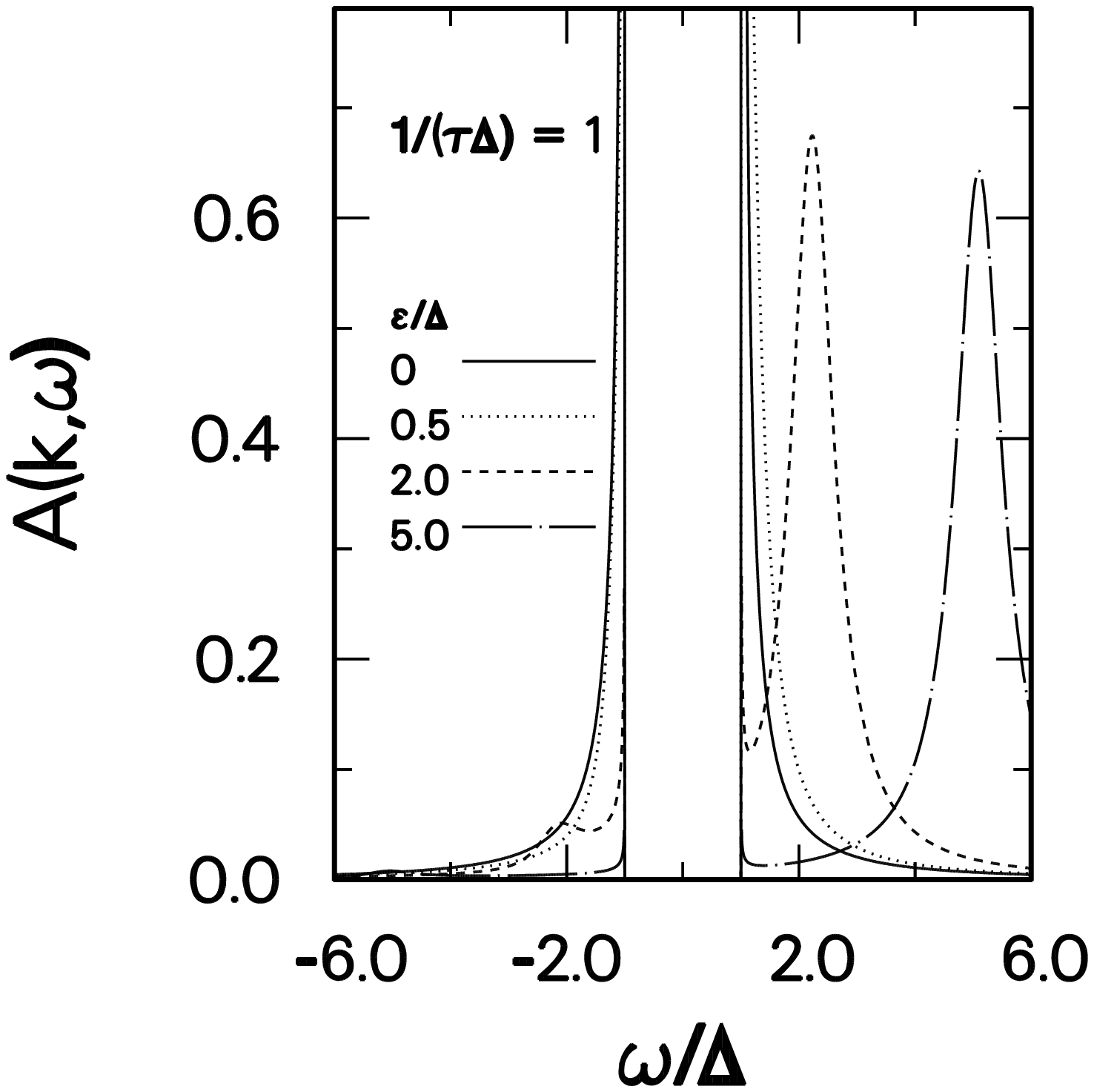,height=7.5in,width=6.7in}
\vskip+0mm
\caption{The single-particle spectral function, $A(k,\omega)$, vs.
normalized frequency, $\omega/\Delta$, with $1/(\tau \Delta) = 1$, for
various quasiparticle energies, $\epsilon/\Delta$. The Fermi surface
result (solid curve) has particle-hole symmetry. The result for 
$\epsilon/\Delta = 0.5$ would have a peak within the gap (between -1 and 1)
except that the coherence factors in Eq. (\protect\ref{greennew}) give zero
residue for the gap region. As the quasiparticle energy increases, the spectral
function begins to resemble the normal state spectral function. Small
square-root singularities still exist, nonetheless at the particle and hole 
gap edges.}
\end{figure}
\vfil\eject

\begin{figure}
\psfig{file=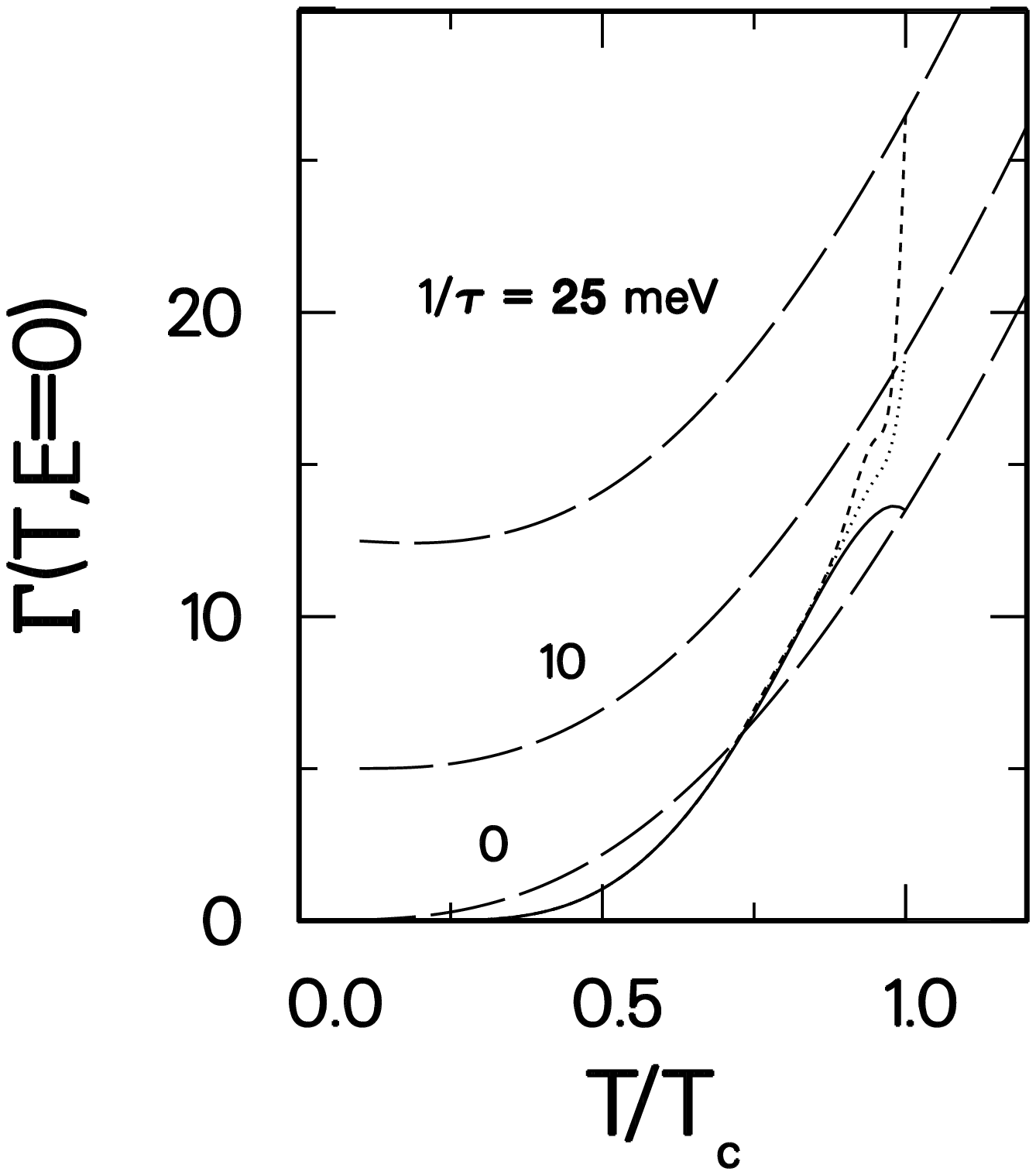,height=7.5in,width=6.7in}
\vskip+0mm
\caption{The quasiparticle
scattering rate, $\Gamma(T,E=0)$ vs. $T/T_c$ for various impurity
scattering rates, $1/\tau$, in both the superconducting and normal
(long-dashed curves) states. In the clean limit (solid curve) there is
an enhancement immediately below $T_c$. When impurity scattering is
present, this scattering is immediately suppressed in the superconducting
state, as shown by the dotted and dashed curves. Below a temperature of
about 0.8$T_c$ the scattering rate is independent of the amount of impurity
scattering present in the normal state. A Debye electron-phonon spectrum
was used with $\lambda = 1$ and cutoff frequency, $\omega_D = 30$ meV.}
\end{figure}
\vfil\eject

\begin{figure}
\psfig{file=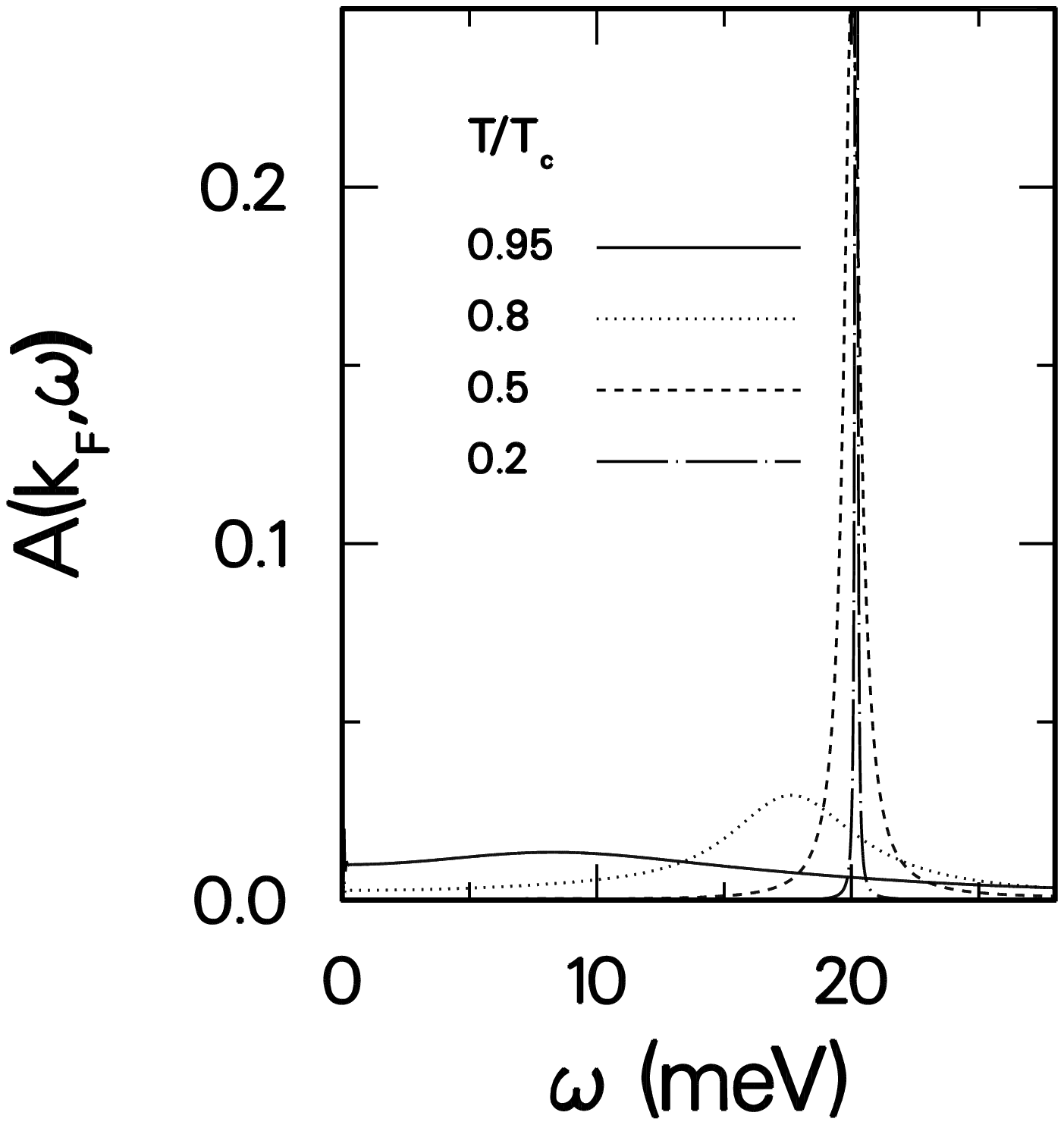,height=7.5in,width=6.7in}
\vskip+0mm
\caption{The single-particle spectral function, $A(k_F,\omega)$ at the
Fermi surface, vs.
frequency, in the clean limit, for various reduced temperatures. The spectral
function is considerably broadened near $T_c$, due to temperature alone. At
the two lowest temperatures shown, while there is no true gap in the excitation
spectrum, this plot makes it clear that, practically speaking, an effective
gap in the excitation spectrum is present. A Debye electron-phonon spectrum
was used with $\lambda = 1$ and cutoff frequency, $\omega_D = 30$ meV.}
\end{figure}
\vfil\eject

\begin{figure}
\psfig{file=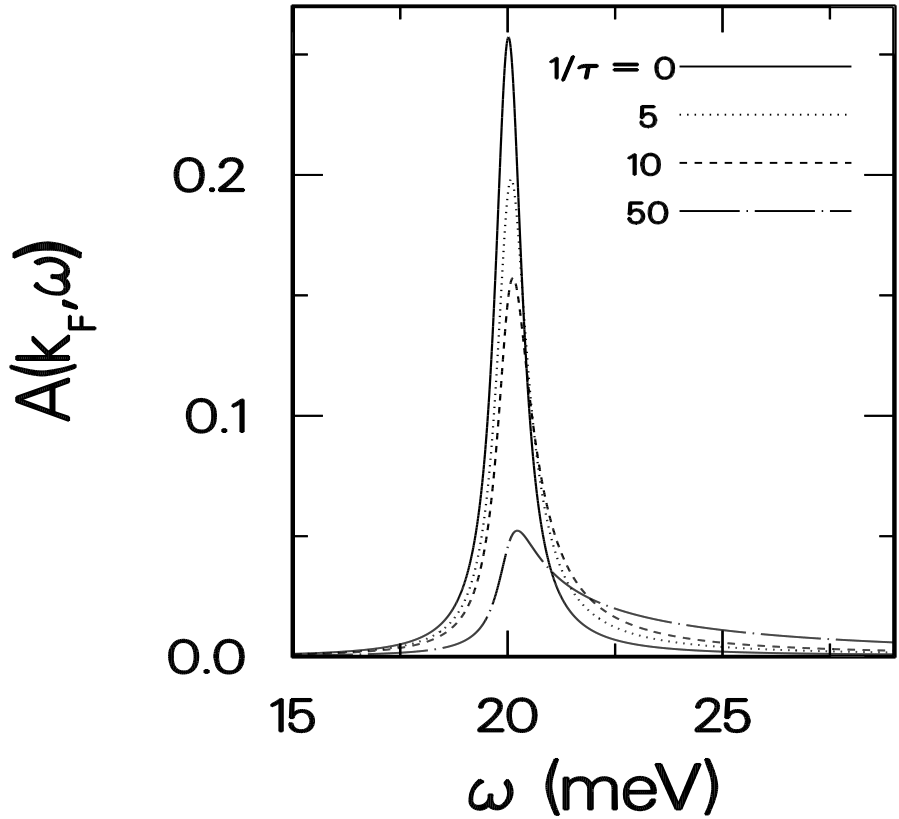,height=7.5in,width=6.7in}
\vskip+0mm
\caption{The single-particle spectral function, $A(k_F,\omega)$ at the
Fermi surface, vs.
frequency, for various impurity scattering rates, as indicated. Note the
broadening which occurs with increasing impurity scattering. However, spectral
weight remains absent in the ``gap region''.
Results are for a temperature $T/T_c = 0.5$, and with a Debye electron-phonon
spectrum with $\lambda = 1$ and cutoff frequency, $\omega_D = 30$ meV.}
\end{figure}
\vfil\eject

\begin{figure}
\psfig{file=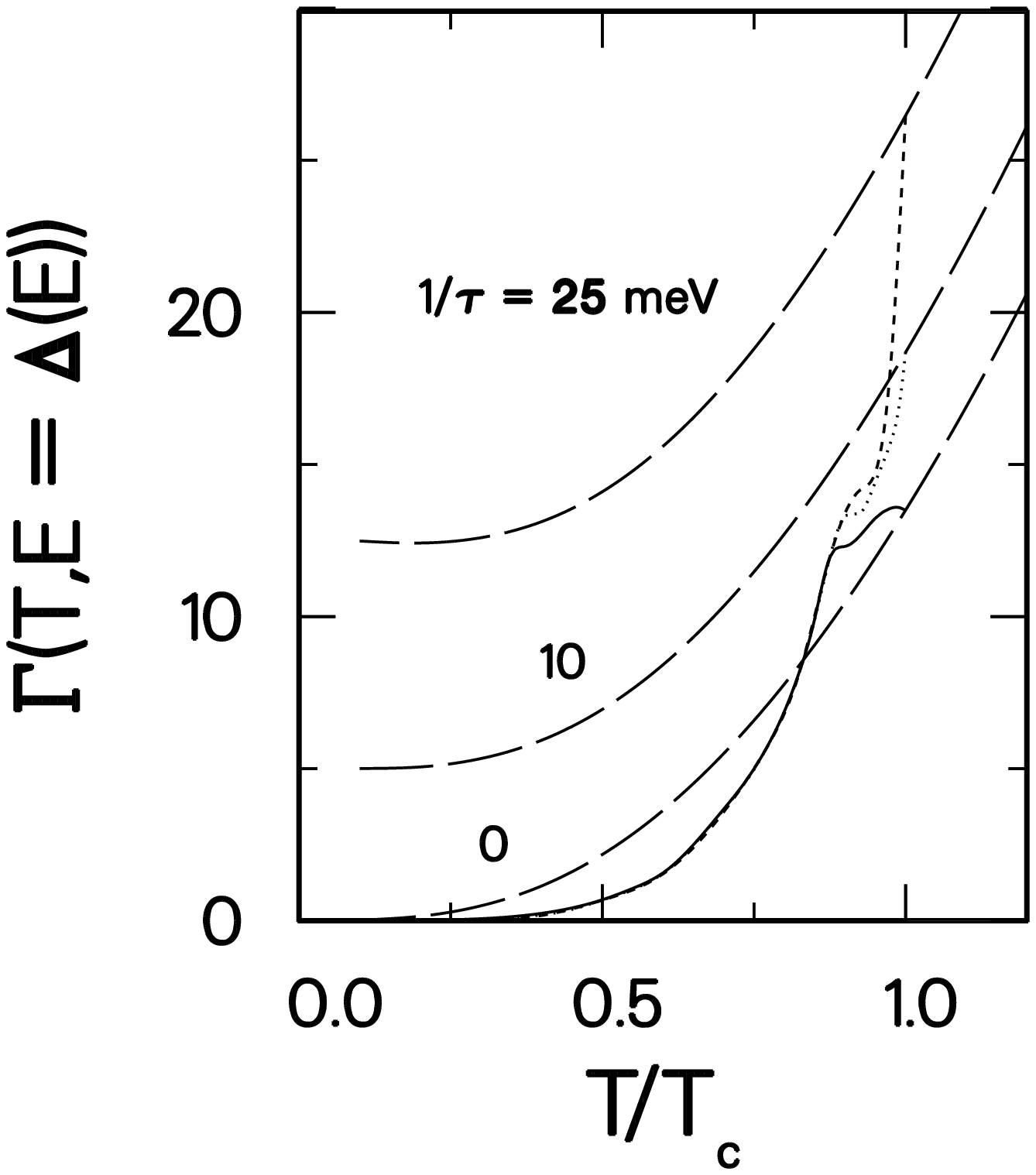,height=7.5in,width=6.7in}
\vskip+0mm
\caption{The quasiparticle
scattering rate, $\Gamma(T,E=\Delta(E))$, evaluated at the quasiparticle
energy, given on the Fermi surface by $E = \Delta(E)$,  vs. $T/T_c$ for 
various impurity scattering rates, $1/\tau$, in both the superconducting 
and normal (long-dashed curves) states. These results are in quantitative
agreement with those in Fig. 6, since the energy scale, $\Delta$, is
still small compared to other (phonon) energy scales in the problem.
In particular, below 
about 0.8$T_c$,  the scattering rate is independent of the amount of impurity
scattering present in the normal state. A Debye electron-phonon spectrum
was used with $\lambda = 1$ and cutoff frequency, $\omega_D = 30$ meV.}
\end{figure}
\vfil\eject

\begin{figure}
\psfig{file=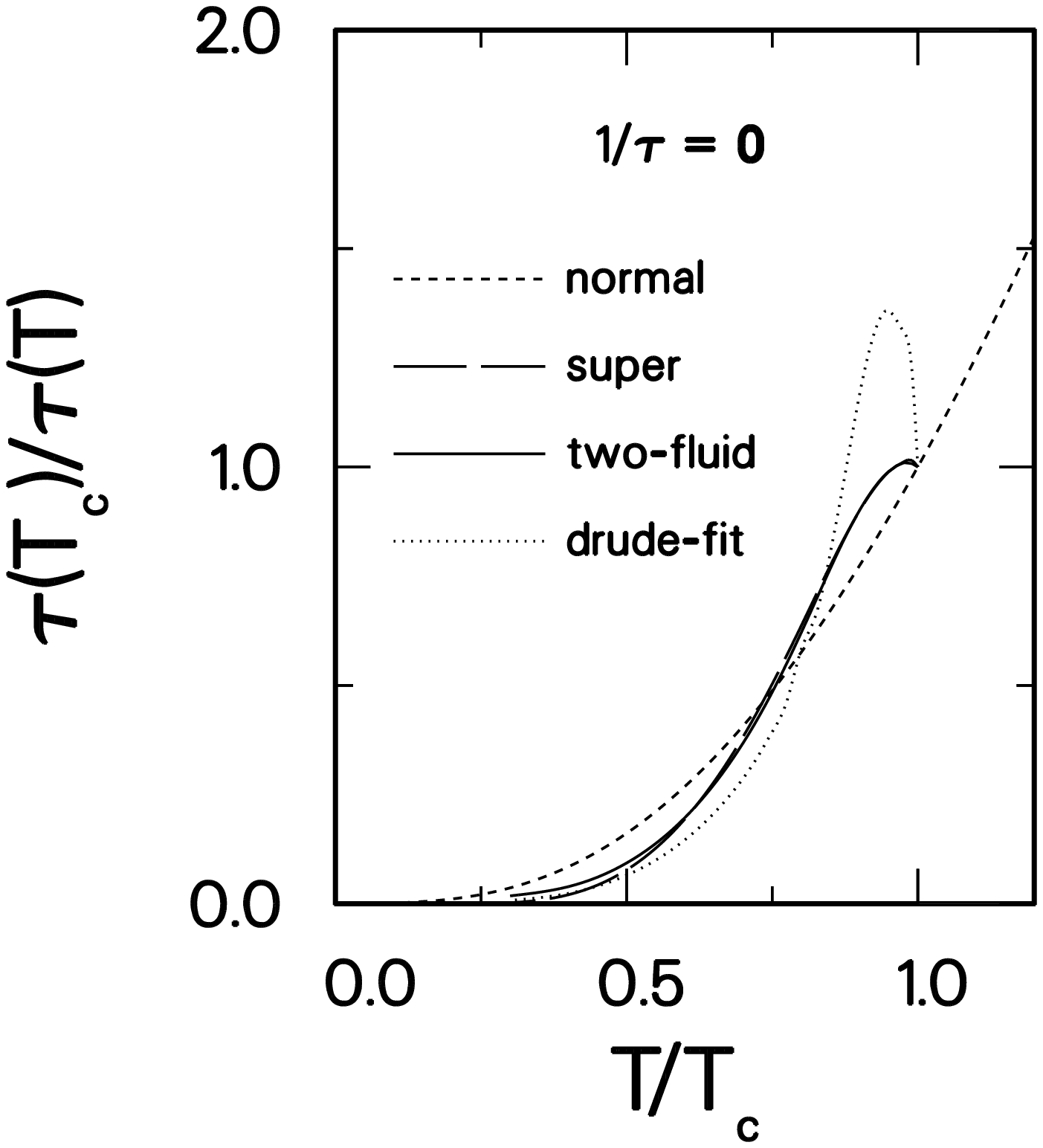,height=7.5in,width=6.7in}
\vskip+0mm
\caption{Various normalized scattering rates vs. reduced temperature.
The normal and superconducting scattering rates come from the quasiparticle
inverse lifetime, given by Eq. (\protect\ref{lifetime}),
at zero energy. The two-fluid
result comes from Eq. (\protect\ref{klein_onet}) while the
Drude fit is obtained by fitting Eq. (\protect\ref{romerofit})
to the low frequency conductivity in the
superconducting state. Note the agreement of the scattering rate as
extracted from the two-fluid analysis with the inverse lifetime in the
superconducting state. A Debye electron-phonon spectrum
was used with $\lambda = 1$ and cutoff frequency, $\omega_D = 30$ meV.}
\end{figure}
\vfil\eject

\begin{figure}
\psfig{file=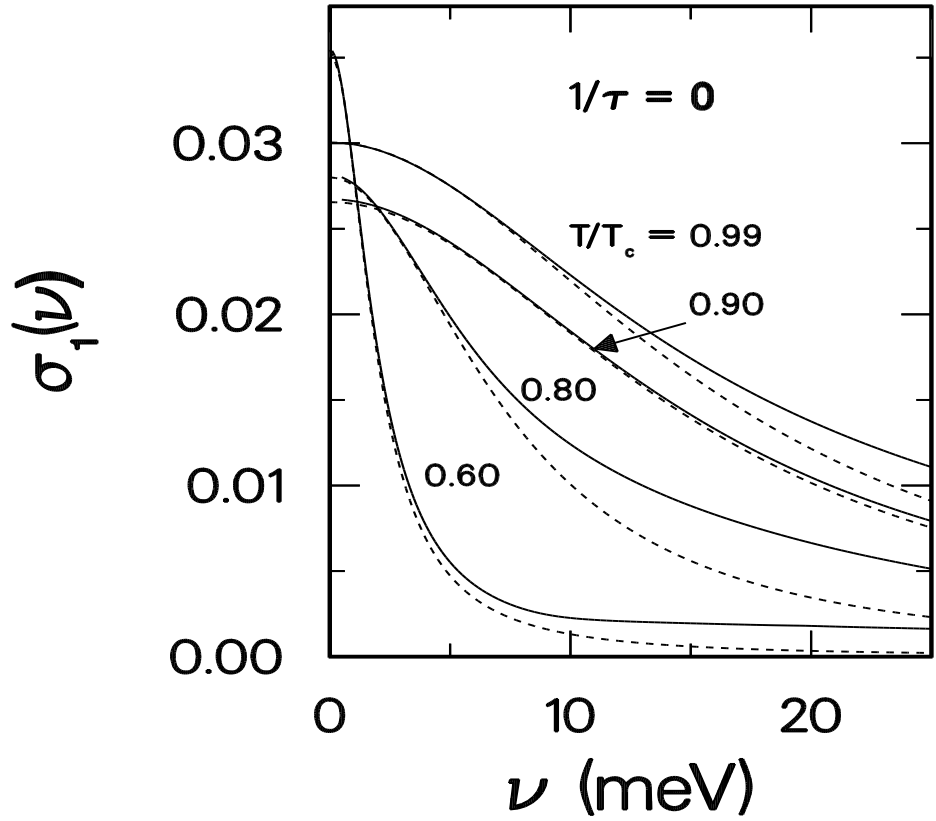,height=7.5in,width=6.7in}
\vskip+0mm
\caption{The low frequency conductivity in the superconducting
state (solid curves) along with their fits based on  Eq.
(\protect\ref{romerofit})
(dashed curves). These fits were used in Fig. 10 (dotted curves).}
\end{figure}
\vfil\eject

\begin{figure}
\psfig{file=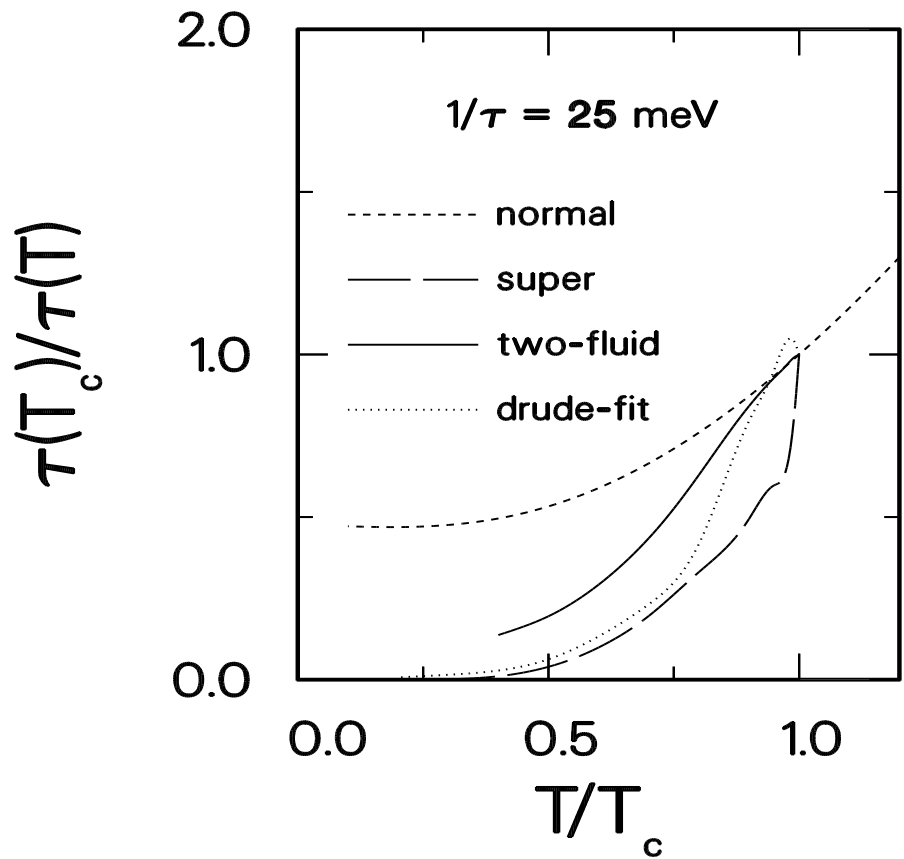,height=7.5in,width=6.7in}
\vskip+0mm
\caption{Same as Fig. 10, except now with an impurity scattering rate
of 25 meV. Note that the two-fluid analysis agrees poorly with the 
quasiparticle inverse lifetime.}
\end{figure}
\vfil\eject

\begin{figure}
\psfig{file=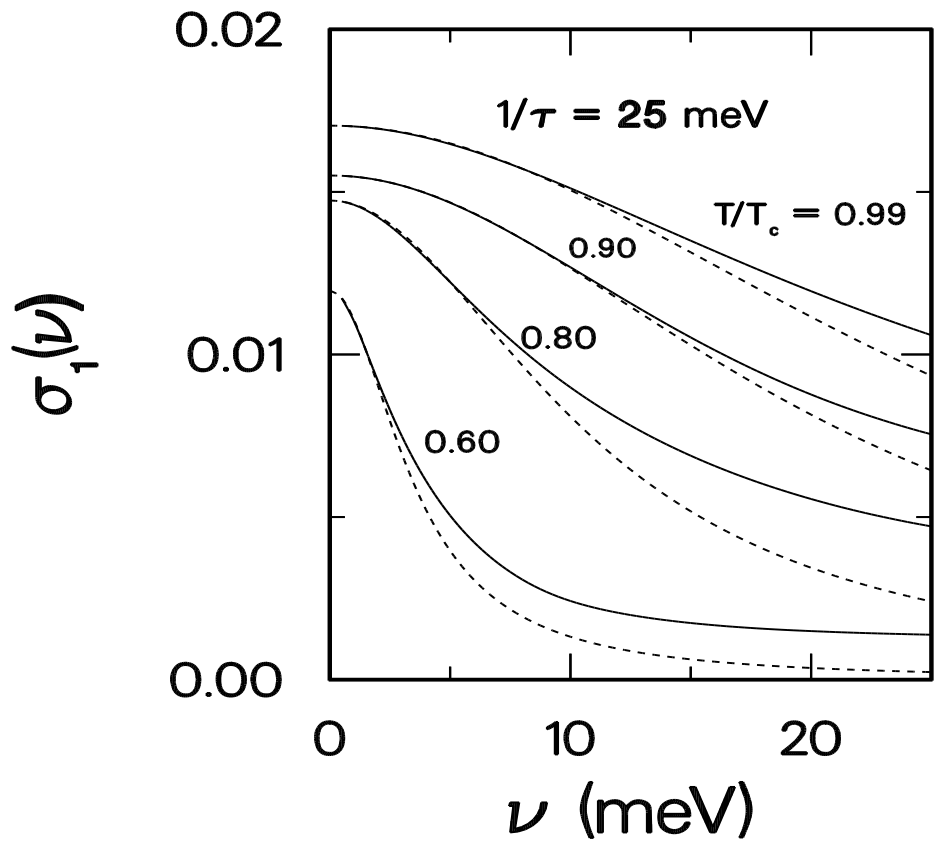,height=7.5in,width=6.7in}
\vskip+0mm
\caption{The low frequency conductivity fits used in Fig. 12.}
\end{figure}
\vfil\eject

\begin{figure}
\psfig{file=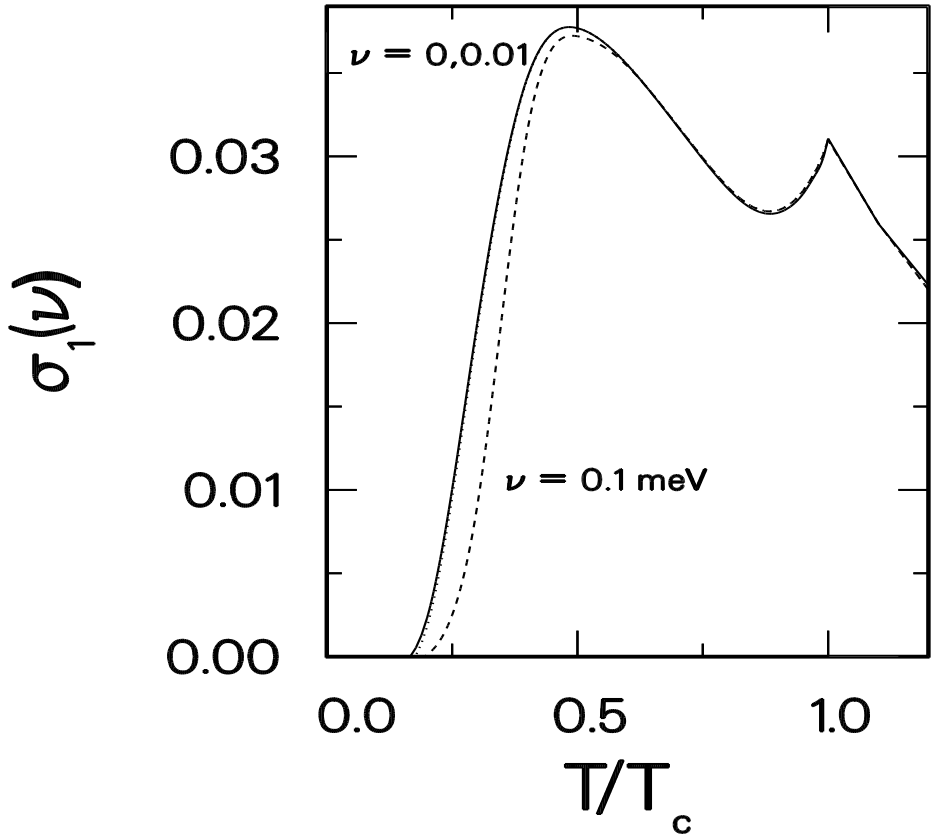,height=7.5in,width=6.7in}
\vskip+0mm
\caption{The very low frequency conductivity as a function of
reduced temperature, in the clean limit.  A Debye electron-phonon spectrum
was used with $\lambda = 1$ and cutoff frequency, $\omega_D = 30$ meV.
Note that the results are relatively  insensitive to frequency (the
$\nu = 0.01$ meV result, given by the dotted curve, is essentially hidden
by the zero frequency result). A microwave
experiment yields essentially  zero frequency results.}
\end{figure}
\vfil\eject

\begin{figure}
\psfig{file=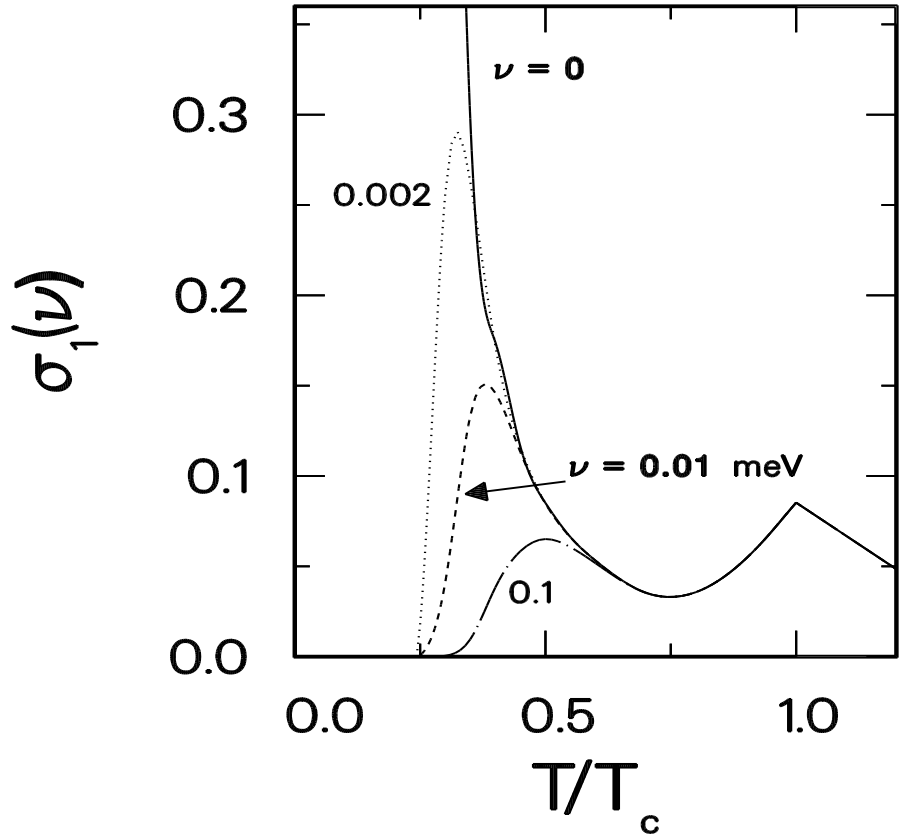,height=7.5in,width=6.7in}
\vskip+0mm
\caption{Same as for Fig. 14, but now with the triangular spectrum
as described in the text. Note that the zero frequency conductivity appears
to diverge as $T \rightarrow 0$. A microwave experiment will yield a large
low temperature peak as a function of reduced temperature, whose magnitude
will depend strongly on the frequency. These results are for the clean limit.
The peak is also reduced as impurity scattering is added (not shown).}
\end{figure}
\vfil\eject

\begin{figure}
\psfig{file=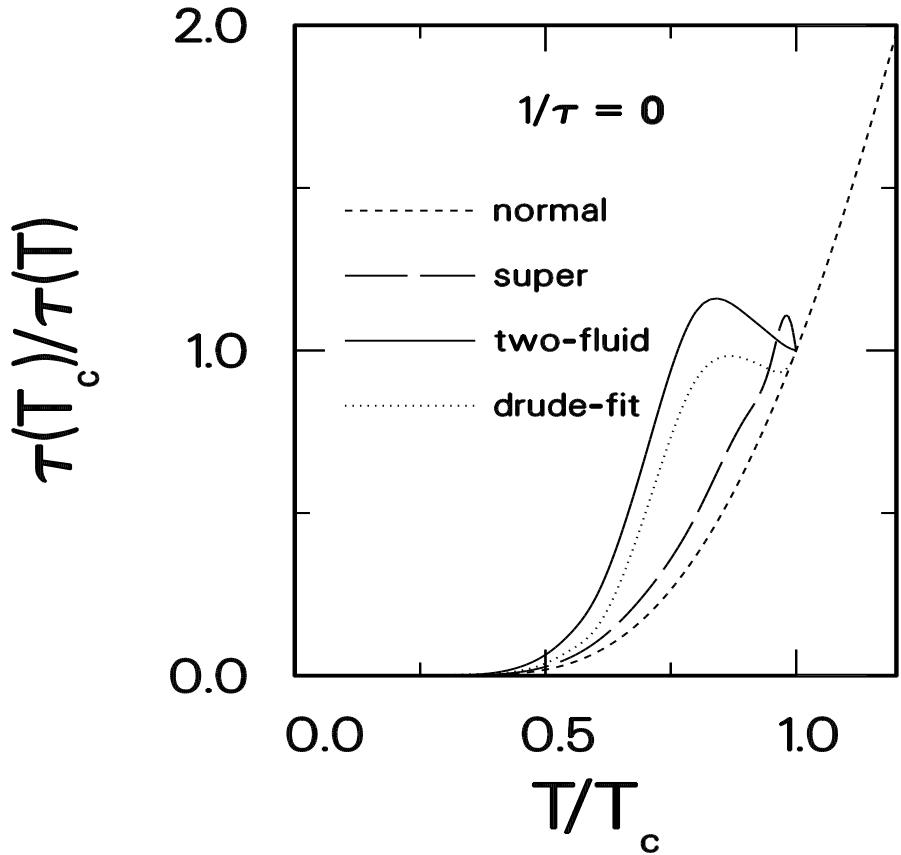,height=7.5in,width=6.7in}
\vskip+0mm
\caption{Comparison of scattering rates (as in Fig. 10) calculated 
for the triangular spectrum, in the clean limit. Note that the normal
state result approaches $T=0$ exponentially due to the gap in the 
$\alpha^2F(\nu)$ spectrum (in Fig. 10 the corresponding curve approached
zero with a power law behaviour). While the results in the superconducting
state are qualitatively similar, they no longer agree quantitatively with
one another, as in Fig. 10.}
\end{figure}
\end{document}